\documentclass[aps,pre,twocolumn,showpacs,superscriptaddress,groupedaddress,flaotfix]{revtex4-2}

\usepackage[caption=false]{subfig}
\usepackage{graphicx} 
\usepackage{mathrsfs}
\usepackage{amsfonts}
\usepackage{amsmath}
\usepackage[noabbrev]{cleveref}
\usepackage[toc,page]{appendix}
\usepackage{import}
\usepackage[nice]{units}
\usepackage{float}
\usepackage{placeins}
\usepackage{color}

\begin{document}
\title{Cumulative geometric frustration and super-extensive energy scaling in non-linear classical XY-spin model.}
\date{\today}
\author{Snir Meiri}
\author{Efi Efrati} 
\email{efi.efrati@weizmann.ac.il} 
\affiliation{Department of
Physics of Complex Systems, Weizmann Institute of Science, Rehovot
76100, Israel}

\begin{abstract}
Geometric frustration results from a discrepancy between the locally favored arrangement of the constituents of a system and the geometry of the embedding space. Geometric frustration can be either non-cumulative, which implies an extensive energy growth, or cumulative which implies super-extensive energy scaling and highly cooperative ground state configurations which may depend on the dimensions of the system. Cumulative geometric frustration was identified in a variety of continuous systems including liquid crystals, filament bundles and molecular crystals. However, a spin-lattice model which clearly demonstrates cumulative geometric frustration was lacking. In this work we describe a non-linear variation of the XY-spin model on a triangular lattice that displays cumulative geometric frustration. The model is studied numerically and analyzed in three distinct parameter regimes, which are associated with different energy minimizing configurations. We show that, despite the difference in the ground state structure in the different regimes, in all cases the super-extensive power-law growth of the frustration energy for small domains grows with the same universal exponent that is predicted from the structure of the underlying compatibility condition.
\end{abstract}
\maketitle

\section{\label{intro}Introduction\protect\\}
Geometric frustration arises whenever the geometry of the constituents of an assembly and their interactions favor a local motif that is incompatible with long range order in the ambient space in which the system is embedded. Because the favored local motifs are ill-fitting for long-range propagation, the ground-state configuration of a given domain cannot be obtained by local minimization of the individual interactions. Instead, the ground state is obtained via a compromise that best resolves the conflict, which might be non-local. In recent years, geometric frustration was shown to underlie the mechanism that leads to a variety of non-trivial morphological responses, including size limitation \cite{Gra20}, filamentation \cite{MPNM14} and other non-trivial response properties \cite{LS16,GSK18,AAMS14,ZGDS19}. Such unique characteristics have been observed in several different settings including a colloidal crystal confined to a spherical interface \cite{MPNM14}, filamentation of irregular elastic hexagons \cite{LW17}, twisted molecular crystals \cite{HAS+18} and bundles of twisted filaments \cite{Gra12}. Recent efforts seek to provide a unified framework that describes geometric frustration in these diverse systems and unravels its common underlying mechanism \cite{Gra16,ME21}.

Perhaps the most renowned geometrically frustrated system, in particular in the field of hard condensed matter, is the Ising anti-ferromagnet on a triangular lattice \cite{Wan50,KN53}. This system is comprised of spins of values $\pm1$ that are located at the vertices of a triangular lattice. The interactions are set such that every pair of adjacent spins favors anti-alignment according to the Hamiltonian $H=J\sum_{\langle i,j \rangle}^{}s_i\cdot s_j ,\quad 0<J$.
Although the minimal energy per interaction is $-J$, the minimal energy per triangular facet which includes three edges is not $-3J$ as one could na\"ively expect but rather $-J$ as well. This is a result of the topology of the lattice, as at least one of the edges in every facet must connect spins of the same state. One can always find a configuration that achieves this lower bound throughout the system, rendering the system's energy extensive. Moreover, the frustration in this system as well as in closely related systems can be integrated out of the system in a  coarse-grained description \cite{RL19}. The frustrated Ising anti-ferromagnet on a triangular lattice does not exhibit any of the exotic morphological response properties described above. Additional types of frustrated spin systems include the XY model frustrated by a uniform magnetic flux through the lattice \cite{TJ83,Hal85,Tei88,LBH+18} and the XY model embedded on hyperbolic surfaces that is frustrated due to the curvature of the embedding space \cite{BSK09}. 
In these works the studied XY spin systems were examined in the large geometric frustration regime, and the observed ground states did not exhibit super-extensive energy scaling nor long-range cooperativity.
The planar formulation of both systems requires choosing a gauge, rendering the comparison of the obtained textures ambiguous. 
The continuum limit of the systems is given by the nematic liquid-crystalline phase on a curved surface, which was shown to yield super-extensive energy scaling on domains small compared to the radius of curvature \cite{NE18,ME21}. In the above mentioned works, the geometric length-scale associated with the frustration was comparable to the size of the lattice unit cell. The resulting frustration saturation is the probable culprit for the suppression of the expected super-extensive energy scaling. In the opposite limit of small frustration we expect the models to also exhibit super extensive energy scaling and highly cooperative ground states, but these regimes are yet to be explored.

A well studied system exhibiting a highly cooperative ground state due to geometric frustration is that of bent-core liquid-crystals \cite{Mey76,HC20,NE18,Sel21}. In this system the liquid crystal forming molecules locally favor an arrangement of vanishing twist and splay and a constant bend $b_0$. Considering the two dimensional case, the system is described in terms of a planar unit vector field that indicates the mean direction of the orientation of the long axis of the molecules in a small volume element, termed the director. The director can be defined as: $\hat{n}=\left( \cos{\theta},\sin{\theta} \right)$ where $\theta$ is the angle the director forms with the x-axis. The perpendicular director field is then defined as: $\hat{n}_{\perp}=\left( -\sin{\theta},\cos{\theta} \right)$.  Using these definitions one can relate the directional derivatives of the director with the splay and signed bend functions, respectively, through:
\begin{equation}
      s=\nabla \cdot \hat{n}= \hat{n}_{\perp} \cdot \nabla \theta  \quad \text{and} \quad b=\hat{n} \cdot \nabla \hat{n} =\hat{n} \cdot \nabla \theta. 
\end{equation}
The simplest liquid crystalline phase is the simple nematic where the director is uniform in space and both $b$ and $s$ vanish in the ground state. The energetic cost of small deviations from this preferred state are given by the Frank-free-energy, which is quadratic in both $s$ and $b$.  
For bent-core liquid crystals the system locally favors a state of vanishing splay and a constant bend giving rise to the modified Frank-free-energy:
\begin{equation}
 H = \frac{1}{2}\int [K_1 s^2+K_3\left(b-b_0\right)^2 ]dA,
\end{equation}
where $K_1$ and $K_3$ are the Frank coefficients related to splay and bend, respectively \cite{Mey76,Sel21,note1}. Since the splay and bend fields are not the native variables of the system, these fields cannot assume arbitrary values. In order to correspond to a viable planar director configuration, these fields must satisfy the compatibility condition \cite{NE18}:
\begin{equation} \label{eq:CompBCLC}
s^2+b^2+\hat{n} \cdot \nabla s - \hat{n}_{\perp} \cdot \nabla b=0,
\end{equation}
From this compatibility condition it is clear that a uniform field of vanishing splay and constant non-vanishing bend cannot be formed, and necessitates the emergence of gradients in these fields. The combination of splay and bend gradients that constitute the ground state configuration depends on the ratio of the Frank coefficients and the dimensions of the system. In the limits of $K_1 \ll K_3$ and $K_1 \gg K_3$ the ground state can be analytically obtained and their energy in isotropic domains of area $A$ was shown to grow as $E \propto A^2$ \cite{NE18}. The case of $K_1 = K_3$, was numerically solved and shown to follow the same energy scaling.
 This value of the energy scaling exponent can be predicted directly from the structure of the compatibility condition \cite{ME21}.
As the director orientation can be measured globally relatively to the $x$ axis and the preferred misalignment is prescribed along the director, no gauge freedom remains in the system.

The geometric frustration of systems whose ground state is highly cooperative and associated with a super-extensive energy (i.e. an energy term that grows faster than the mass of the system) was termed cumulative geometric frustration \cite{HG21,ME21}. 
The frustration in systems whose ground states are associated with extensive energy scaling and with short-range or no cooperativity, is termed non-cumulative frustration. The Ising anti-ferromagnet, as well as a variety of frustrated binary spin models were recently shown to support only non-cumulative frustration \cite{RL19}.

To the best of the knowledge of the authors no spin lattice model was shown to exhibit super-extensive energy scaling due to frustration. As spin-lattice models are important tools for establishing theoretical understanding of complex physical behavior our main aim in what follows is to formulate and study a model that will display cumulative geometric frustration, and super-extensive energy scaling with a continuum limit that is well understood. We next come to construct such a model inspired by planar bent-core liquid crystals. The model demonstrates super-extensive energy scaling, long-range cooperativity and distinct conformations that best resolve the frustration in different parameters regimes. Despite the variance in the ground state textures for the different parameters, we show that the power law of the energy scaling is universal and can be predicted from 
the continuum limit of the corresponding compatibility condition.

\section{\label{model}The model\protect\\}
\begin{figure}[t]
\includegraphics[width=4.3cm]{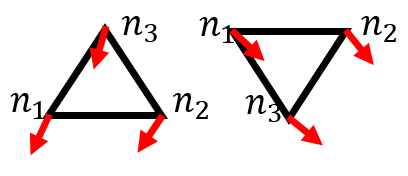}
\caption{  Upright (left) and up-side down (right) triangles' notations for vertices.}
\label{fig:Notations_1}
\end{figure}

The constituents of the model are classical XY spins located at the vertices of a triangular lattice. 
Unlike the interactions in ferromagnetism and antiferromagnetism that tend to align or anti-align the spins, respectively, the spin interactions in our model are designed to favor only slightly mis-aligned orientations. 
The designed  non-linear interactions are inspired by bent core liquid crystals. Each spin is characterized by its relative angle to the x-axis, $\theta\in [ 0,2\pi)$. The orientations of the three spins at each triangular facet are assumed to not be too dissimilar, such that the average direction, $\bar\theta=(\theta_1+\theta_2+\theta_3)/3$, is well-defined and not far from the vertices' $\theta_i$ values. We next define on every facet two auxiliary variables:
\begin{equation} \label{eq:4}
\begin{split}
\Psi_1=&-\frac{\theta_2-\theta_1}{l}\sin(\bar{\theta})\pm
\frac{2\theta_3-\theta_2-\theta_1}{\sqrt{3}l}\cos(\bar{\theta}),\\
\Psi_2=&\frac{\theta_2-\theta_1}{l}\cos(\bar{\theta})\pm
\frac{2\theta_3-\theta_2-\theta_1}{\sqrt{3}l}\sin(\bar{\theta}),
\end{split}
\end{equation}
where the $+$(-) signs above corresponds to upright (upside down) triangles according to the notations defined in figure \ref{fig:Notations_1} and $l$ is the length of edges in the lattice. 
The Hamiltonian expressed in terms of the auxiliary variables reads:
\begin{equation}
\mathcal{H}=\frac{1}{2}\sum_{\text{facets}}K_1\Psi_1^2+K_3(\Psi_2-b_0)^2.\label{Hamiltonian}
\end{equation}

To unravel the source of frustration in this system, one should first scrutinize the structure of the compatibility condition associated with the auxiliary variables $\Psi_i$. The Hamiltonian involves only two auxiliary variables per facet, $\Psi_1$  and $\Psi_2$, which are naturally insufficient to fully determine all three spin values of the vertices of the facet.  
Given the values of $\Psi_1$  and $\Psi_2$, the space of possible spin values reduces to a one-dimensional curve in the space of all possible three spin values at the vertices. Considering two adjacent facets that share an edge we  obtain four $\Psi$ values that, in turn, fully determine the associated four spin values. Each facet in the bulk belongs to three distinct edge-sharing facet-pairs, in which the local $\Psi$ values determine all spin values. The compatibility condition emerges from the requirement that the spin values calculated through each of these three distinct pairs, agree. 
Considering the continuum limit where the edge length $l\to 0$, associating finite differences with gradients, identifying $\Psi_1=s$ as the splay and $\Psi_2=b$ as the signed bend, and expanding the values of $s$ and $b$ around their values at the central facet yields, to first order in $l$, the known compatibility condition: $s^2+b^2+\hat{n} \cdot \nabla s - \hat{n}_{\perp} \cdot \nabla b=0$ \cite{NE18}. For more details see appendix \ref{app:Banana}. Similarly to the case of bent-core liquid crystals, this compatibility condition precludes the formation of a configuration of vanishing $\Psi_1$ and constant non-vanishing $\Psi_2$, thus preventing setting the energy in \eqref{Hamiltonian} to vanish globally. 

As was recently suggested in \cite{ME21}, the structure of the compatibility conditions encodes information about the ground state solutions for isotropic small enough domains compared to the length-scale associated with the frustration. For such small domains, the energy, which is approximately quadratic in the generalized strains, $\Psi-\bar{\Psi}$, can be expanded in orders of the spatial coordinates. Each successive order includes higher spatial derivatives of the generalized strains and contributes less to the total energy.  As a result one may follow a na\"ive approach to construct explicit solutions by setting the energy to vanish order-by-ordern (for more details see \cite{ME21}). 
In the present case, the resulting compatibility condition in the continuum limit yields a first order differential relation in the generalized stains. Thus one may set the zero order of the generalized strains to zero, yet must incorporate a
combination of a gradient of splay in the direction of the director and/or a gradient of bend in the perpendicular direction, resulting in a super-extensive energy scaling of $E\propto A^2$. See appendix \ref{app:local}.

\begin{figure*}[!t]
\includegraphics[width=16.2cm]{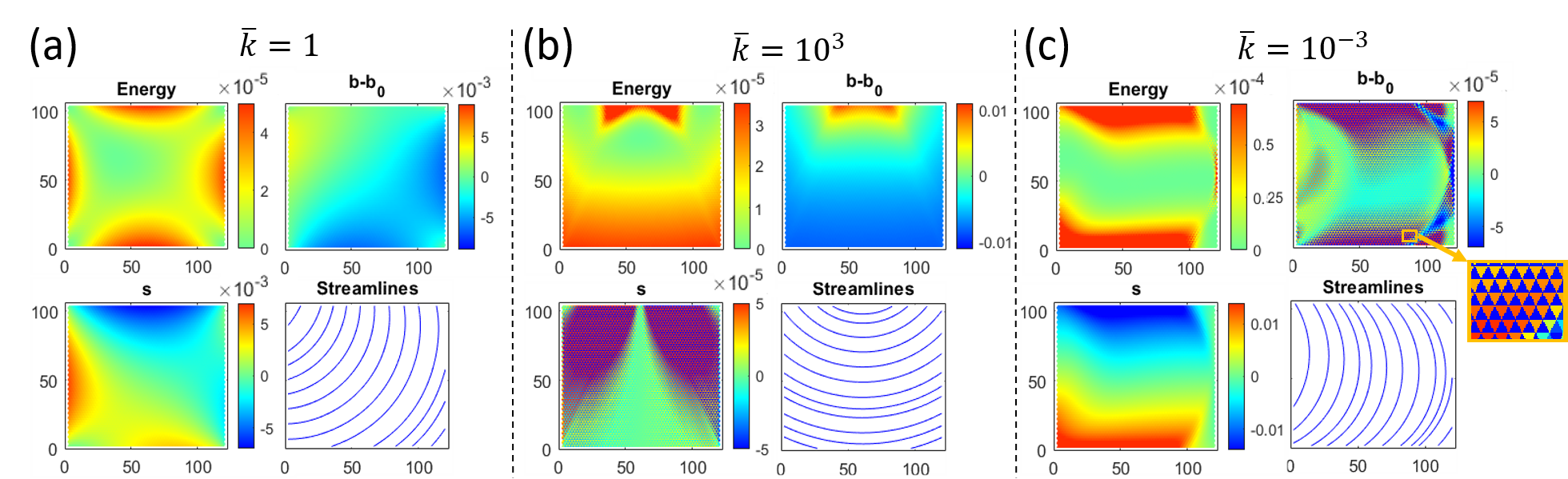}
\caption{ Results of lattices of size $60\times60$ for different limiting values of $\overline{k}$: (a) $\overline{k}=1$, (b) $\overline{k}=10^{3}$  and (c) $\overline{k}=10^{-3}$ for $b_0=0.015$. For every quartet color-maps of energy (top-left), bend-$b_{0}$ (top-right) and splay (bottom-left) are presented together with an interpolated plot of the resulting streamlines (bottom-right). The inset on the right displays a magnification of the area marked in the orange rectangle in $b-b_{0}$ in (c). Note that the panels employ different color-bar scales.}
\label{fig:Examp}
\end{figure*}

\section{\label{results}Results\protect\\}
\begin{figure}[h !]
\includegraphics[width=5.7cm]{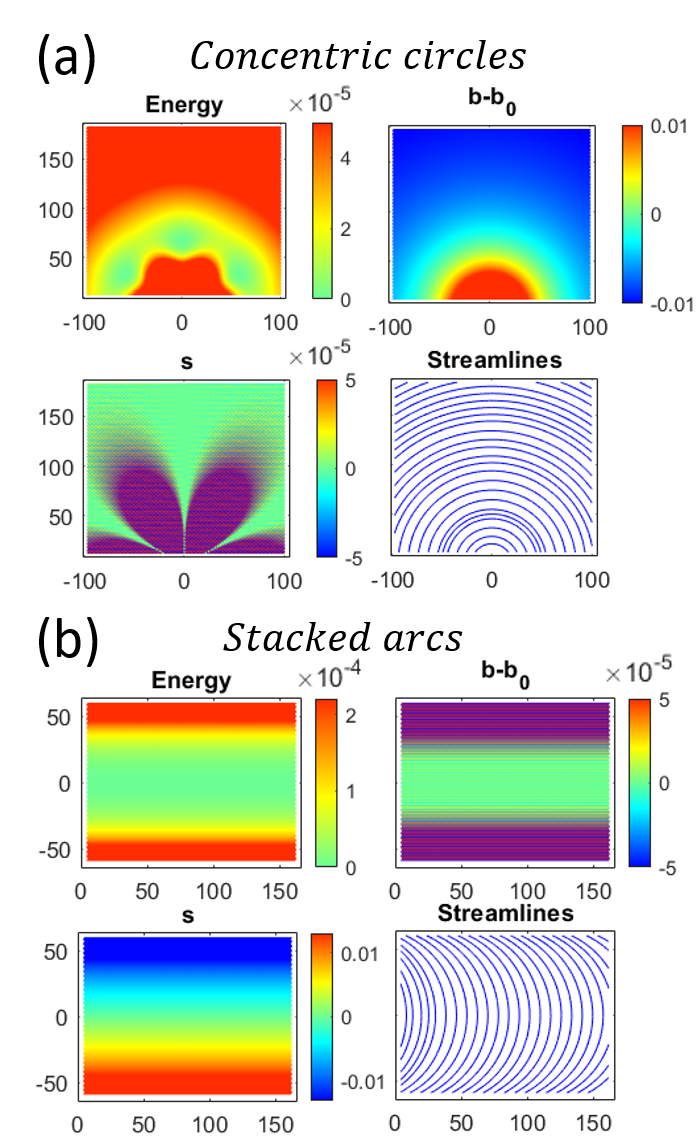}
\caption{ Lattices with spin values sampled from the continuous solutions of vanishing splay (a) and constant bend (b) for $b_0=0.015$. The lattice sizes are $ 100 \times 100 $ sites with the origin of the circles located at (0,-10) in (a) and $ 70 \times 80 $ sites with the symmetry line along the x axis in (b). For every quartet color-maps of energy (top-left), bend-$b_{0}$ (top-right) and splay (bottom-left) are shown together with an interpolated plot of the streamlines (bottom-right). Note that the panels employ different color-bar scales.}
\label{fig:ContSamp}
\end{figure}

\begin{figure*}[t !]
\includegraphics[width=13.3cm]{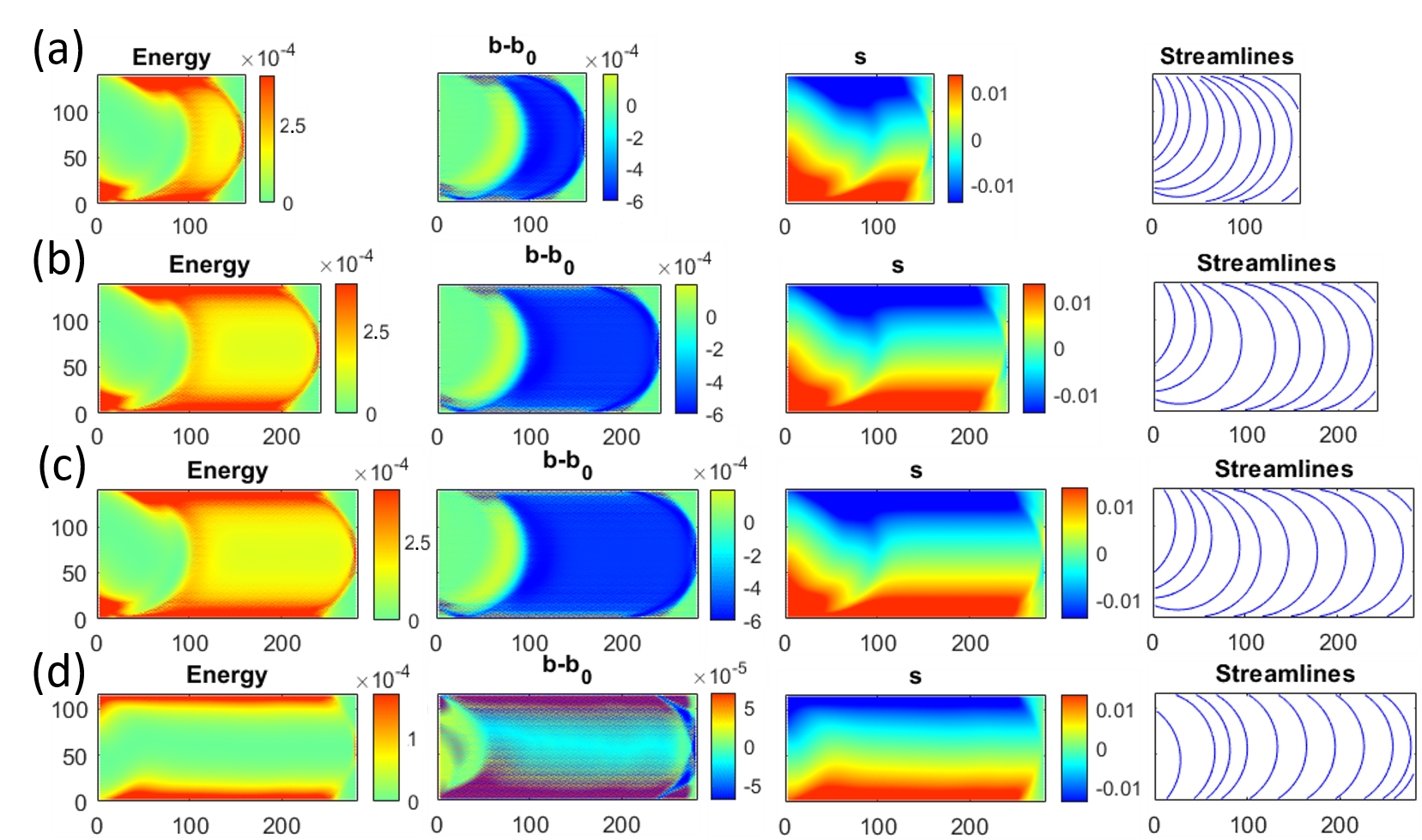}
\caption{ Results of ribbon-like lattices for $\bar{k}=10^{-3}$ and $b_0=0.015$. (a)-(d) Lattices of sizes $80\times 80$, $120\times 80$, $140\times 80$ and $140\times 66$ sites, respectively. 
From left to right, the plots in each line show the energy, bend-$b_{0}$, splay and an interpolated plot of the resulting streamlines. Note that the panel employ different color-bar scales.}
\label{fig:Ribbon}
\end{figure*}

\begin{figure*}[!tp]
\includegraphics[width=17.9cm]{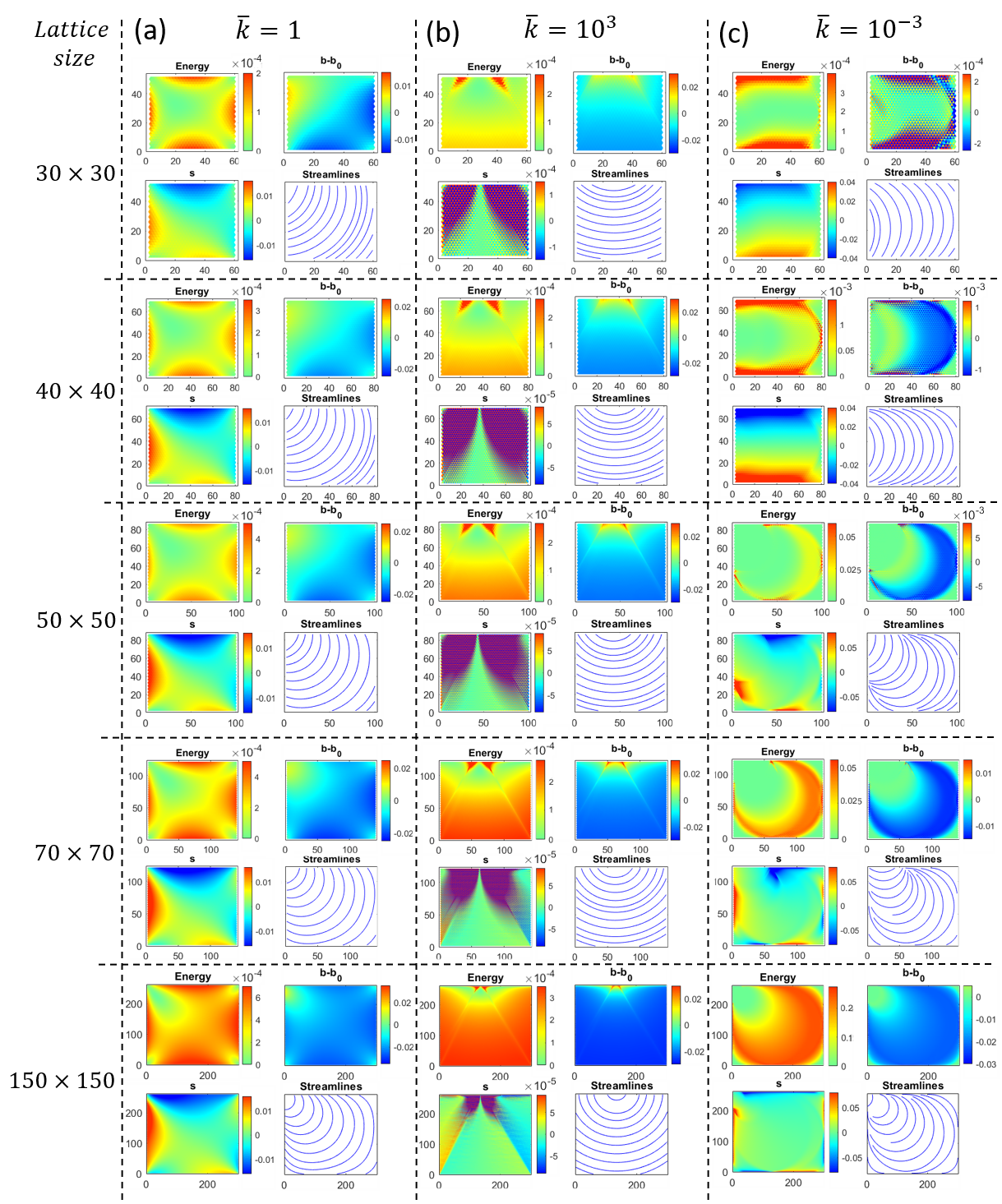}
\caption{  Results of isotropic lattices of different sizes stated on the left for different limits: (a) $\overline{k}=1$, (b) $\overline{k}=10^{3}$  and (c) $\overline{k}=10^{-3}$ for $b_0=0.03$. For every quartet color-maps for energy (top-left), bend-$b_{0}$ (top-right) and splay (bottom-left) are presented together with an interpolated plot of the resulting streamlines (bottom-right).}
\label{fig:examp_0_03}
\end{figure*}

The model was studied numerically by minimizing the Hamiltonian to find the zero temperature ground state at different values of $\bar{k}=K_1/K_3$. The resulting lowest energy states display non-uniform deviations of the variables $\Psi$ from their locally desired values. The parameters chosen for the first part (figures \ref{fig:Examp}-\ref{fig:Ribbon}) are $\left( \bar{\Psi}_1,\bar{\Psi}_2 \right) =\left( 0,0.015 \right) $ and for the second part (figure \ref{fig:examp_0_03}) $\left( \bar{\Psi}_1,\bar{\Psi}_2 \right) =\left( 0,0.03 \right) $. In all case $l=2$, and for the isotropic domains (Figures \ref{fig:Examp} and \ref{fig:examp_0_03}) domains with equal number of spins along the two edges are chosen, i.e. $N_x=N_y=N$. The resulting side lengths thus measure $L_x=2 N$ and $L_y=\sqrt{3}N$. Free boundary conditions are used throughout this work.

It is straightforward to show that the locally favored values of the auxiliary functions cannot be globally obtained and thus lead to geometric frustration.
The numerically obtained ground state configurations are of highly cooperative nature and assume very distinct formations for the three studied values of $\bar{k}$, as can be seen for domains of $60 \times 60$ sites that are shown in figure \ref{fig:Examp} (for additional results see SM).

For small enough domains for the case of $\bar{k}=1$ the deviations of the splay and bend each show a single hourglass form, displaying maximal deviations at the centers of opposite edges, yet oriented along perpendicular directions. The energy distribution naturally exhibits a crossed hourglasses form with peaks at the centers of each of the domain's four edges (see SM).
As the domain size is increased the 
symmetry axis of the streamlines of the spin field drifts away from the mid-line of the domain and points more toward its corners, as can be seen in figure \ref{fig:Examp} (a). Subsequently, the splay and bend variations lose their hourglass form. Yet,
the energy distribution keeps the crossed hourglass form even for large domains.

Panels (b)-(c) in Figure \ref{fig:Examp} show the results of the splay-dominated and bend-dominated ground-states resulting from $\bar{k}=10^3$ and $\bar{k}=10^{-3}$, respectively. The energy as well as the splay and bend deviations in these two cases differ substantially compared to the former results. In particular, the splay deviations for $\bar{k}=10^3$ and the bend deviations for  $\bar{k}=10^{-3}$ display staggered textures with upright and upside down triangles displaying deviations of opposite signs. This manifests in the purple-like regions in the respective plots, as can also be seen in the magnified region in panel (c). Despite the violent oscillations, the energy distribution in both cases appear to be smooth at least to leading order. 

In order to better understand these results we compare them to the solutions of vanishing splay and constant bend of the corresponding limiting continuous field theory \cite{NE18}. The respective solutions for the limits of $\bar{k}\to\infty$ and $\bar{k}\to 0$ are formations of concentric circles ($s=0$) and stacked identical arcs ($b=b_0$). To compare our lattice system with the known continuous results we consider these two smooth solutions, and inherit from them the local spin orientation. That is, the vertex $i,j$ located at $\left(x(i,j),y(i,j)\right)$ inherits $\theta_{ij}=\theta\left(x(i,j),y(i,j)\right)$. For more details regarding the sampling procedure, see appendices \ref{app:Concentric} and \ref{app:arcs}. The corresponding fields resulting from this sampling procedure are shown in figure \ref{fig:ContSamp}. As can be seen in panel (a), although the director field the solution is sampled from is axially symmetric, the resulting splay field displays a six-fold symmetry (only top half is shown). The splay deviations show a staggered texture with six-fold azimuthal periodicity as well as nontrivial radial dependence. 

The resulting values of bend and splay shown here can be examined by positioning the center of a single upright (upside down) triangle in the location $\left( r,\theta \right)$, where $r$ is the distance from the origin of the concentric circles and $\theta$ is the relative angle the line connecting the two creates with the x-axis. Sampling the spins from the director and expanding the resulting bend and splay values to second order in $l/r$, where $l$ is the facet edge length, results in:
 \begin{align}
    b_{up,down}&\simeq\frac{1}{r} \pm \frac{l \sin{\theta}}{2 \sqrt{3} r^2}, \\ s_{up,down}&\simeq \pm \frac{l \cos{\theta}\left( 1-2 \cos{2 \theta}\right)}{2 \sqrt{3} r^2}.
\end{align}
 As can be seen from this expansion, the bend field is smooth to leading order while the leading order in splay results in opposite signs of deviations for inverted triangles with magnitude that decays as $1/r^2$ and has a six-fold symmetry, as seen in figure \ref{fig:ContSamp} (a). Carrying out the same procedure in the case of the stacked arcs where the symmetry line of the arcs consolidate with the x-axis, and expanding to leading orders of the dimensionless parameter $l\cdot b_0$ results in:
 \begin{align}
 b_{up,down}&\simeq b_0 \mp \frac{l b_0^3 y}{4 \sqrt{3} \left(1-b_0^2 y^2 \right)^2}, \\ s_{up,down}&\simeq b_0^2 y \frac{-1+b_0^2 y \left(y\mp \frac{l}{4 \sqrt{3}}  \right)}{\left(1-b_0^2 y^2 \right)^{3/2}}. \label{eq:supdown} 
\end{align}

While for the case of $\bar{k}=10^{3}$ the results seem to be in good agreement with the results of the sampling procedure, up to the discussed discontinuities and some minor deviations, the results for $\bar{k}=10^{-3}$ show more prominent discrepancies, mostly localized near the boundaries of the domain. While the ideal stacked arcs formation is inherently symmetric along the x-axis, the configurations obtained through numerical minimization break this symmetry, mainly for lattices of larger size. The most prominent effects are located at the regions near the edges oriented perpendicular to the symmetry axis of the stacked arcs formation (right and left sides in Figure \ref{fig:Examp} (c)). The continuous reorientation associated with a constant bend phase causes the splay of the director field to increase with the distance from the mid-line of the domain. On the right hand side of the domain the middle region of the stacked arcs is outside the domain. This allows a rigid reorientation of the arcs portions that fill the corners of the domain. This reorientation, in turn, significantly reduces the energy associate with the splay. On the left portion of the domain the stacked arcs also reorient. This boundary feature is not entirely understood, yet seems to enlarge the region of low energy in the domain.

The above arguments, that describe the local reorientation and the structure of the boundary layer perturbation to the stacked arcs phase expected in the $\bar{k}\to 0$ limit, assumes that such a phase with $b=b_0$ everywhere exists and is associated with a finite (albeit possibly high) splay energy. This is not always the case, as is made evident by considering a domain of width $W> 2/b_0$. Such domains exhibit a divergence of the splay for the $b=b_0$ solution, as seen in \eqref{eq:supdown}. To avoid this attempted divergence, the bend is reduced below $b_0$ pushing the splay singularity further away. This behavior can observed in figure \ref{fig:Ribbon}.
The total energy in this case is dominated by the bend variations introduced to mitigate the attempted splay singularity. The boundary term variations that locally reduce the splay energy that we discussed and observed for narrower domains also occur here, and are localized to the regions near the right and left boundaries of the domain that do not increase in size as we consider longer ribbons, see Figure \ref{fig:Ribbon}.

In order to study the behavior of domains of spatial length-scale, $L_y$, larger than the geometrical length scale associated with the reference curvature, $b_0$, we studied the case of $b_0=0.03$ as well. This allows us to probe the behavior in this regime while keeping a moderate number of studied sites. A representative sample of the resulting minimal configurations in the three studied limits is displayed in Figure \ref{fig:examp_0_03}. 

In the case of large domains global aspects of the solutions, such as the size of the domain and its aspect ratio become prominent.
Any non-trivial resolution of frustration necessitates the introduction of spatial gradients in $\Psi_1$ and $\Psi_2$ defined in \eqref{eq:4}. In solutions of large domains these gradients no longer remain uniform and vary in magnitude, orientation or both due to global considerations. 
Such variations include for example the reorientation of the principal gradient direction visible in the transition between $N=40$ and $N=50$ in figure \ref{fig:examp_0_03}(c).

The form that equation \eqref{eq:CompBCLC} assumes for $b=b_0$ predicts a divergence of $s$ within a finite distance from the solutions' symmetry axis. The distance to this singularity scales linearly with $1/b_0$. Thus, to circumvent this attempted singularity in domains whose width is larger than $2/b_0$, the attempted uniform bend solution (in which $b=b_0$) will center about a constant bend value that is smaller than $b_0$ corresponding to less curved arcs. Such partial straightening is visible between $N=30$ and $N=40$ of Figure \ref{fig:examp_0_03} (c). Similarly, 
the $s=0$ attempted solutions visible in \ref{fig:examp_0_03} (b) originally display concentric arcs distributed symmetrically around the radius $1/b_0$ ($N$=30). However, as the center (focus) of these concentric circle is associated with divergent bending, as the domains increase in size the arc distribution no longer remains symmetric and we observe an increasing abundance of arcs of radii larger than $1/b_0$, as observed for $40\le N$. 


We recall that the only uniform solution is the nematic texture of vanishing bend and splay. In all three studied cases of $\bar{k}$ the solution in the bulk seems to approach this trivial nematic phase, with the saturation energy per unit area of $\varepsilon_\infty =\frac{1}{2}K_3 {b_0}^2$. Nevertheless, in each case of $\bar{k}$ the system approaches this "trivialization" in a different manner, as can be seen in figure \ref{fig:examp_0_03}.


\begin{figure}[h!]
\includegraphics[width=8.5cm]{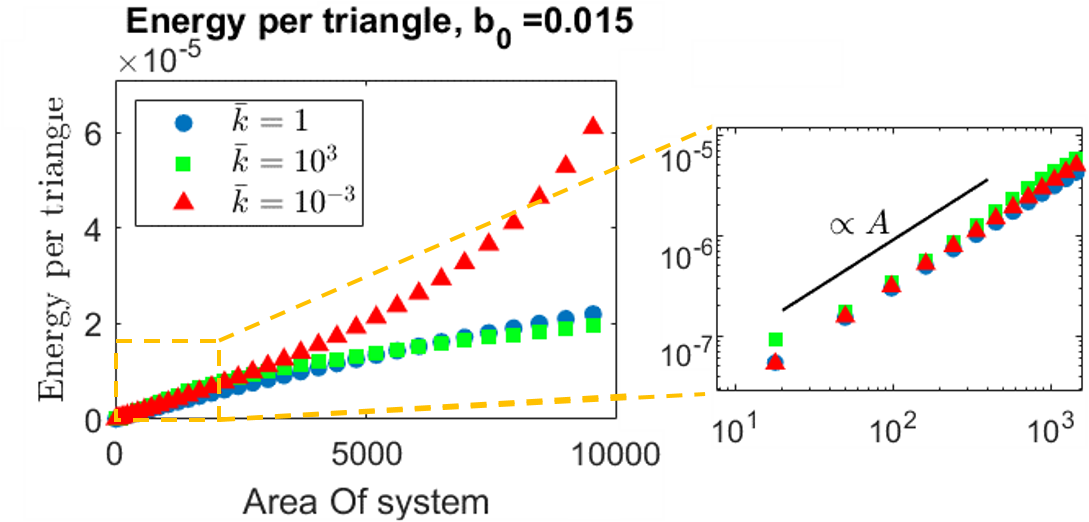}
\caption{ The energy per triangle vs. the area of the system (number of triangles in the lattice) for domains of equal number of sites in the length and in the width. The results are for three limiting values of \(\bar{k}=K_1/K_3\). The results of \(\bar{k}=1\), \(\bar{k}=10^3\) and \(\bar{k}=10^{-3}\) are marked as blue circles, green squares and red triangles, respectively. The scaled plot to the right shows the region near the origin in log-log scale. The black line denotes a linear scaling for guidance. }
\label{fig:Energy}
\end{figure}

\begin{figure}[h!]
\includegraphics[width=6.5cm]{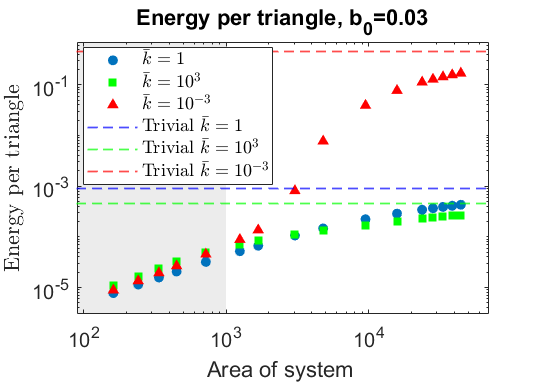}
\caption{ The energy per triangle vs. the area of the system (number of triangles in the lattice) for domains of equal number of sites along the length and along the width of the domain, in logarithmic scale for \(b_0=0.03\). The results presented are for three limits of \(\bar{k}=K_1/K_3\). The results for $\bar{k}=1$ $(K_1=K_3=2)$, $\bar{k}=10^3$ $(K_1=10^3 , K_3=1)$, and $\bar{k}=10^{-3}$ $(K_1=1 , K_3=10^3)$, are marked as blue circles, green squares and red triangles, respectively.
The energy cost of the trivial nematic solution is $\frac{K_3}{2} {b_0}^2$. As the different values of $\bar{k}$ correspond to different values of $K_3$, the asymptotes of the energy vary accordingly.}
\label{fig:Energy2}
\end{figure}

While the ground state solutions for the three values of $\bar{k}$ differ substantially from one another it has been recently shown that the energy associated with such frustrated solutions in small enough domains depends primarily on the associated compatibility conditions \cite{ME21}. In the present case the structure of the compatibility condition implies the universal power law $E\propto M^2$ where $M$ is the mass of the domain (or equivalently that the energy per unit area scales linearly with the area). As the domains grow in size higher order effects lead to deviations from this power-law as can be observed in figure \ref{fig:Energy} and Figure \ref{fig:Energy2}. For example, for the case of $\bar{k}=10^{-3}$, in both figures, as the systems' dimensions reach the internal length-scale $1/b_0$ the attempted strain singularity causes a dramatic increase in the energy growth rate. Figure \ref{fig:Energy2} also shows that the three studied systems associated with the different limits of $\bar{k}$ all approach the trivial nematic solution corresponding to vanishing splay and bend as their area grows. This mechanism of frustration saturation is not unique. In particular, one might expect the formation of defects absorbing the frustration as the systems grow in size. Within each of the separated defect-free-domains the solution may be associated with smaller generalized strains and thus with lower energy. 
Such solutions are favorable when energy decrease facilitated by the incorporation of defects is greater than the cost associated with the defect formation. In the case presented here the use of the signed bend strongly penalizes common defected textures associated with bent core-liquid crystals, such as layered textures and hexagonal defect formations \cite{Sel21}.

Nonetheless, we note that in the shaded region, denoting domains whose dimensions are smaller than $1/b_0$, we observe in all cases the same universal scaling behaviour.

\section{\label{discussion}Discussion\protect\\ }
In this manuscript we present a spin-lattice model that demonstrates cumulative geometric frustration which results in cooperative ground state configurations that depend on global attributes of the domain. The model presented here was studied in three different constitutive parameter regimes, captured by the parameter $\bar{k}$. While the energy minimizing configurations for the three cases differ substantially from one another, in all cases the energy grows as the area squared (for sufficiently small isotropic domains). This universal exponent is in turn predicted from the structure of the compatibility condition which determines the form of the optimal compromise of the frustration. Although systems of different $\bar{k}$ agree on the energy scaling for small domains, the energy scaling differs for larger domains until it plateaus for systems much larger than the geometric length-scale of the system. The values at which the energy saturates coincide with the energetic cost of the nematic trivial solution for the different $\bar{k}$.

In case of a unit cell smaller than the internal geometric length-scale, yet comparable in magnitude, frustration saturation is reached in small domains of a few unit cells. Cumulative frustration would only be evident within this narrow region of domain sizes. Stronger frustration corresponds to smaller geometric length-scales, which may become smaller than the dimensions of a single unit cell. In such cases one might not detect any signature of the cumulative nature of the frustration, and the system would display extensive energy scaling.

In this work we focused on the properties of the ground states at $T=0$, primarily in isotropically growing domains, demonstrating the super-extensive energy scaling due to geometric frustration. In more realistic models, additional thermodynamic considerations, such as accounting for surface tension and entropy, are required to adequately describe a given system. Such an approach, predicting the structural fate of an assembly was recently proposed \cite{HG21}. Cumulative geometric frustration was shown to potentially lead to filamentation. In the present setting this would imply that the length of a domain $L$ becomes much larger than its width $W$. 

The question whether a given system would have the tendency to form filamentous domains and the regime at which this tendency exists depends on the manner in which its energy approaches the frustration saturation energy $\varepsilon_\infty$. In cases where the approach to $\varepsilon_\infty$ is slow enough, growth arrest along the $W$ dimension, leading to ribbons of any desired width, depending on the thermodynamic parameters, can be achieved. For faster approach to $\varepsilon_\infty$, such growth arrest can only be achieved up to a finite width, after which only bulk conformations are attainable. Assuming saturation approach of the form $W^{-\nu}$, $0<\nu<1$ corresponds to the first (slow) type while $1<\nu$ corresponds to the latter (fast) \cite{HG21}. In the present case we numerically found $\frac{E}{A} \to A^{-\nu}$, with $0.5 \le \nu \le 0.65$, implying a slow approach to the frustration saturation.  

Many naturally found self-assembled geometrically frustrated structures, such as protein assemblies, can only support a discrete conformation space. This in turn leads to discrete values of the gradients, which may change the nature of the compatibility conditions and the associated optimal compromises. Within the context of the present model this corresponds to quantizing the allowed orientations of the spins, similarly to the $n$-states Potts model.
For the limit of $n\to\infty$ one expects to recover the behavior of the continuous case described above. For moderate values of $n$ one can expect rich and possibly different response properties that depend on the particular value of $n$ which determines the granularity of the response.

\begin{acknowledgments}
This work was funded by the Israel Science Foundation Grant No. 1479/16 and  Grant No. 1444/21. EE thanks the Ascher Foundation for their support. 
\end{acknowledgments}

\appendix

\section{\label{app:Concentric} Sampling from the concentric circles solution}
To sample the solution of concentric circles we employ a lattice of size $100 \times 100$ sites with the lower edge aligned with the x-axis and centered compared to it. Next, the spin values are sampled from the continuous concentric circle texture by orienting the spins along the tangent to the continuous director field at the vertex location. By setting the center of the concentric circles at (0,-10) the direction of the director at each point reads: $\theta \left(x,y\right)= \tan^{-1} \left(\frac{y+10}{x}\right) + \pi/2.$

\section{ Sampling from the stacked arcs solution \label{app:arcs}} 

The solution of stacked arcs is composed of arcs of radius $r_0=1/b_0$ that are perpendicular to a straight line that connects their centers. This line is chosen to coincide with the x-axis. The sampling procedure is done via setting a lattice of size $70 \times 80$ sites such that the flat edges of the lattice are perpendicular to the x-axis and the lattice is centered about the y-axis. Next, the spin values are sampled from the continuous solution by orienting the spins along the tangent to the continuous director field at the vertex location.The director is independent of the x coordinate and reads: $\theta \left(y\right)= \tan^{-1} \left(\frac{y}{\sqrt{r_0^2-y^2}}\right) + \pi/2.$

\section{Compatibility conditions of bent core spins on a triangular lattice} \label{app:Banana}
\begin{figure}[h]
\includegraphics[width=8.0cm]{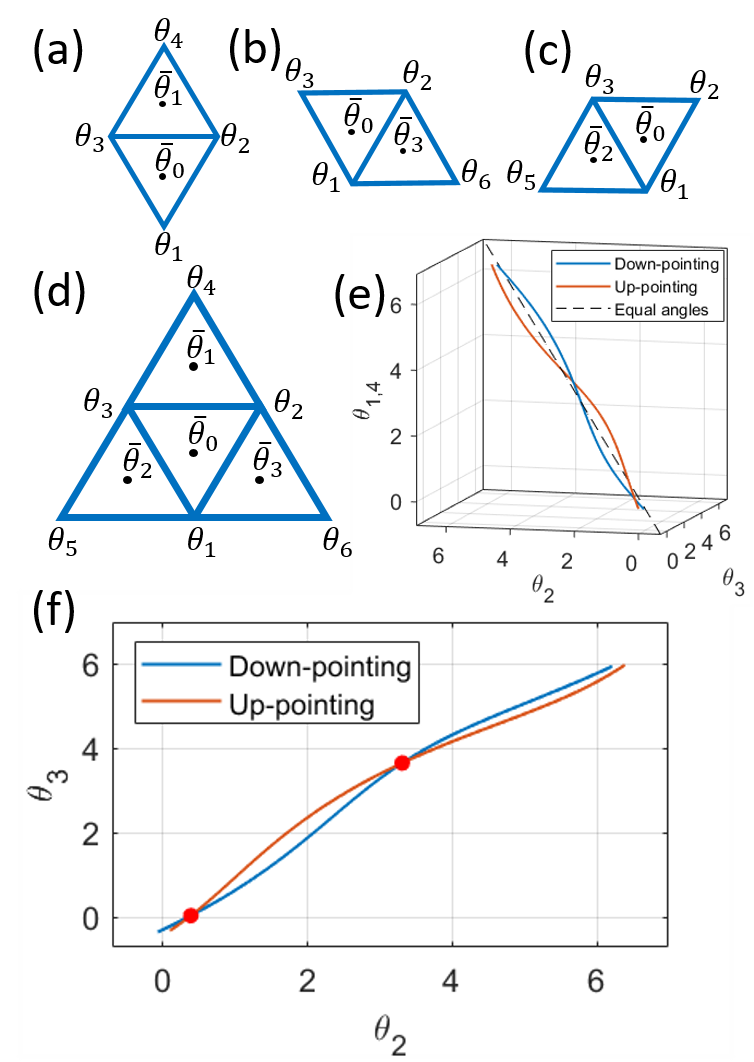}
\caption{ (a)-(d) Notations for compatibility conditions. (a)-(c) Three types of pairs of facets in the lattice sharing an edge. (d) A quartet of connected facets in the lattice. (e)-(f) Graphic solutions for an upright pair for a choice of $b_0=0.3$, $s_0=-0.2$, $b_1=0.4$ and $s_1=0.15$. The blue curve marks allowed spin orientations considering the down-pointing triangle alone and the orange curve is obtained by considering the up-pointing triangle alone. The three dimensional curves are shown in (e), where
black diagonal dashed line marks equal orientations in all the spins. (f) The projections onto the $\{ \theta_2, \theta_3 \}$ plane; crossing points marked in red.}
\label{fig:comp}
\end{figure}

A single facet in the lattice has three spin degrees of freedom located at its vertices. The splay and bend associated with the facet can be directly deduced from the vertices' spin values 
Using equations \eqref{eq:4}. The converse is , however, under-determined; given prescribed values for the bend and splay of a given facet there is a one dimensional space of solutions that corresponds to different choices of spins, as can be seen in figure \ref{fig:comp} (e), which displays two such curves that were computed numerically for certain choices of variables in an up-pointing and a down-pointing triangles. 
A pair of adjacent facets as in figure \ref{fig:comp} (a), is associated with four distinct spin values and two pairs of intrinsic fields associated with the splay and bend of each of the facets. In this system the latter fields fully determine the spin values.
The under-determinacy present for a single facet is eliminated in facet pairs by the requirement that the resulting shared spin values match. This can be considered as the crossing of the projections of the curves onto the plane of the shared spins. As can be seen in figure \ref{fig:comp} (f), more than one such crossing might exist.

The spin values at the vertices of every two adjacent facets are fully determined by the splay and bend values of both facets. However, every single facet participates in three distinct such pairs, as seen in figure \ref{fig:comp} (a-c). The values ascribed to the spins of a facet obtained from different pairs should match, leading to a non-trivial relation between the splay and bend of the three facets. 
This over-determinacy settles with the observation that such triplet have five spin degrees of freedom that are transformed to six values of bend and splay. Three such triplets exists for every bulk element as seen in figure \ref{fig:comp} (d). Only two independent such relations exists per such element, due to transitive relations. 

We next consider a continuum treatment for this bulk element. The locations of all the vertices in figure \ref{fig:comp} when setting the location of $\overline{\theta}_0$ to the origin of the frame of reference are given by: 
\begin{eqnarray}\nonumber
    \Vec{r}_1&=&(0,-\frac{l}{\sqrt{3}}),\quad  \Vec{r}_2=(\frac{l}{2},\frac{l}{2\sqrt{3}}),\quad \Vec{r}_3=(-\frac{l}{2}\frac{l}{2\sqrt{3}}),\\ \nonumber
\Vec{r}_4&=&(0,\frac{2 l}{\sqrt{3}}),\quad \Vec{r}_5=(-l,-\frac{l}{\sqrt{3}}), \quad \Vec{r}_6=(l,-\frac{l}{\sqrt{3}}),\\ \nonumber  \Vec{r}_7&=&(0,\frac{l}{\sqrt{3}}),\quad \Vec{r}_8=(-\frac{l}{2},-\frac{l}{2\sqrt{3}}), \quad  \Vec{r}_9=(\frac{l}{2},-\frac{l}{2\sqrt{3}}).
\end{eqnarray}
Indices 1-6 above refer to the vertices and indices 7, 8 and 9 refer to the locations of $\overline{\theta}_1$, $\overline{\theta}_2$ and $\overline{\theta}_3$, respectively.

 The definitions of bend and splay are invertible and result in: $\partial_x\overline{\theta}=b \cos{\overline{\theta} }-s \sin{\overline{\theta} }$ and $\partial_y\overline{\theta}=b \sin{\overline{\theta} }+s \cos{\overline{\theta} }$. The line integral of the gradient of $\theta$ for every closed loop must vanish. This condition results in:

 \begin{equation} \label{eq:compBulk}
\begin{split}
0=& (\Vec{r}_3-\Vec{r}_1)\cdot\begin{pmatrix} \partial_x\overline{\theta}_3 \\ \partial_y\overline{\theta}_3  \end{pmatrix}+ (\Vec{r}_2-\Vec{r}_3)\cdot\begin{pmatrix} \partial_x\overline{\theta}_2 \\ \partial_y\overline{\theta}_2  \end{pmatrix}+\\+&(\Vec{r}_1-\Vec{r}_2)\cdot\begin{pmatrix} \partial_x\overline{\theta}_1 \\ \partial_y\overline{\theta}_1  \end{pmatrix}
\end{split}
\end{equation}

In the continuum limit where l is small and the change in the spin directions is also small, one can expand the mean directions, splays and bends around their values at the location of $\overline{\theta}_0$, by assigning gradients in bend, splay and the mean direction. Substituting these expanded terms into equation \ref{eq:compBulk} and expanding the expression in orders of l, the edge length, agrees to first order with the compatibility condition of planar director fields, shown in equation \ref{eq:CompBCLC}.

\section{The na\"ive local solutions} \label{app:local}
In order to study the na\"ive local solutions in the three studied limits one should first scrutinize the resulting compatibility condition (for more details see \cite{ME21}). The compatibility condition determines the allowed local conformations. Such conformation associated with the lowest energy cost is the na\"ive local solution. The corresponding compatibility condition in the continuum limit reads:
\begin{equation} \label{eq:CompApp}
s^2+b^2+\hat{n} \cdot \nabla s - \hat{n}_{\perp} \cdot \nabla b=0.
\end{equation}
Expanding the generalized strains, $\varepsilon_s=s$, and $\varepsilon_b=b-b_0$, in  orders of the spatial coordinates yields:
\begin{align}
\varepsilon_s=\varepsilon_s^{0,0}+\varepsilon_s^{1,0}x+\varepsilon_s^{0,1}y+...\quad,\\ 
\varepsilon_b=\varepsilon_b^{0,0}+\varepsilon_b^{1,0}x+\varepsilon_b^{0,1}y+... \quad .
\end{align}
Substituting the expanded expressions in the compatibility condition yields to zeroth order:
\begin{equation} 
\left(\varepsilon_s^{0,0}\right)^2+\left(b_0+\varepsilon_b^{0,0}\right)^2+\varepsilon_{s}^{n} - \varepsilon_{b}^{n_\perp}=0,
\label{eq:compExpanded}
\end{equation}
where $\varepsilon_{s}^{n}=n_x\varepsilon_{s}^{1,0}+n_y\varepsilon_{s}^{0,1}$ denotes the 
oriented first derivative of the splay strain along $\hat{n}$, and 
$\varepsilon_{b}^{n_\perp}=n_{\perp,x}\varepsilon{b}^{1,0}+n_{\perp,y}\varepsilon_{b}^{0,1}$ denotes the oriented first derivative of the bend strain along $\hat{n}_\perp$. There are many possibilities for the values of $\varepsilon_s^{0,0},\varepsilon_b^{0,0},\varepsilon_s^{n}$ and $\varepsilon_b^{n_\perp}$ that satisfy \eqref{eq:compExpanded}. Determining which of these values minimizes the total energy depends on the form of the Hamiltonian of the system and the domain considered. For small enough isotropic domains the lowest order in the strain expansion contribute significantly more than the higher orders. For such domains we seek solutions that satisfy $\varepsilon_s^{0,0}=\varepsilon_b^{0,0}=0$ reducing the compatibility condition to 
\begin{equation} 
\left(b_0\right)^2+\varepsilon_{s}^{n} - \varepsilon_{b}^{n_\perp}=0
\rightarrow
\varepsilon_{b}^{n_\perp}=
\left(b_0\right)^2+\varepsilon_{s}^{n} 
\label{eq:compExpanded2}
\end{equation}
Distributing the gradient values between the splay and bend depends on the energetic cost of each in the Hamiltonian. Assuming the integration is performed over a small enough isotropic domain, of length-scale $l$, where $\hat{n}$ and $\hat{n}_\perp$ can be considered to be oriented along a constant direction, one gets:
\begin{equation} 
\begin{split}
E \propto K_1 \left(\varepsilon_{s}^{n}\right)^2 l^4+K_3\left(\varepsilon_{b}^{n_\perp}\right)^2 l^4=\\A^2 [(K_1+K_3) \left(\varepsilon_{s}^{n}\right)^2 +K_3 (b_0^4+2b_0^2 \varepsilon_{s}^{n})],    
\end{split}
\end{equation}
where $A$ is the area of the domain. The gradients minimizing this energy are:
\begin{equation} 
\varepsilon_{s}^{n}=-b_0^2\frac{K_3}{K_1+K_3},\quad \varepsilon_{b}^{n_\perp}=b_0^2\frac{K_1}{K_1+K_3},
\end{equation}
yielding the energy:
\begin{equation} 
E=\frac{1}{6} A^2 b_0^4 \frac{K_1 K_3}{K_1 + K_3}.
\end{equation}
For $\bar{k}=1$ values of $K_1=K_3=2$ were chosen such that in all three studied values of parameters the na\"ive local solutions would have approximately the same energetic cost.

\bibliographystyle{unsrt}
\bibliography{main.bib}

\begin{thebibliography}{10}

\bibitem{Gra20}
Gregory~M. Grason.
\newblock Chiral and achiral mechanisms of self-limiting assembly of twisted
  bundles.
\newblock {\em Soft Matter}, 16(4):1102--1116, January 2020.

\bibitem{MPNM14}
Guangnan Meng, Jayson Paulose, David~R. Nelson, and Vinothan~N. Manoharan.
\newblock Elastic {{Instability}} of a {{Crystal Growing}} on a {{Curved
  Surface}}.
\newblock {\em Science}, 343(6171):634--637, February 2014.

\bibitem{LS16}
Ido Levin and Eran Sharon.
\newblock Anomalously {{Soft Non}}-{{Euclidean Springs}}.
\newblock {\em Physical Review Letters}, 116(3):035502, January 2016.

\bibitem{GSK18}
Doron Grossman, Eran Sharon, and Eytan Katzav.
\newblock Shape and fluctuations of positively curved ribbons.
\newblock {\em Physical Review E}, 98(2):022502, August 2018.

\bibitem{AAMS14}
Shahaf Armon, Hillel Aharoni, Michael Moshe, and Eran Sharon.
\newblock Shape selection in chiral ribbons: From seed pods to supramolecular
  assemblies.
\newblock {\em Soft Matter}, 10(16):2733--2740, 2014.

\bibitem{ZGDS19}
Mingming Zhang, Doron Grossman, Dganit Danino, and Eran Sharon.
\newblock Shape and fluctuations of frustrated self-assembled nano ribbons.
\newblock {\em Nature Communications}, 10(1):3565, August 2019.

\bibitem{LW17}
Martin Lenz and Thomas~A. Witten.
\newblock Geometrical frustration yields fibre formation in self-assembly.
\newblock {\em Nature Physics}, 13(11):1100--1104, November 2017.

\bibitem{HAS+18}
Asaf Haddad, Hillel Aharoni, Eran Sharon, Alexander~G. Shtukenberg, Bart Kahr,
  and Efi Efrati.
\newblock Twist renormalization in molecular crystals driven by geometric
  frustration.
\newblock {\em Soft Matter}, 15(1):116--126, 2018.

\bibitem{Gra12}
Gregory~M. Grason.
\newblock Defects in crystalline packings of twisted filament bundles. {{I}}.
  {{Continuum}} theory of disclinations.
\newblock {\em Physical Review E}, 85(3):031603, March 2012.

\bibitem{Gra16}
Gregory~M. Grason.
\newblock Perspective: {{Geometrically}} frustrated assemblies.
\newblock {\em The Journal of Chemical Physics}, 145(11):110901, September
  2016.

\bibitem{ME21}
Snir Meiri and Efi Efrati.
\newblock Cumulative geometric frustration in physical assemblies.
\newblock {\em Physical Review E}, 104(5):054601, November 2021.

\bibitem{Wan50}
G.~H. Wannier.
\newblock Antiferromagnetism. {{The Triangular Ising Net}}.
\newblock {\em Physical Review}, 79(2):357--364, July 1950.

\bibitem{KN53}
Kenzi Kan{\^o} and Shigeo Naya.
\newblock Antiferromagnetism. {{The Kagom\'e Ising Net}}.
\newblock {\em Progress of Theoretical Physics}, 10(2):158--172, August 1953.

\bibitem{RL19}
Pierre Ronceray and Bruno Le~Floch.
\newblock Range of geometrical frustration in lattice spin models.
\newblock {\em Physical Review E}, 100(5):052150, November 2019.

\bibitem{TJ83}
S.~Teitel and C.~Jayaprakash.
\newblock Josephson-{{Junction Arrays}} in {{Transverse Magnetic Fields}}.
\newblock {\em Physical Review Letters}, 51(21):1999--2002, November 1983.

\bibitem{Hal85}
Thomas~C. Halsey.
\newblock Josephson-junction arrays in transverse magnetic fields: {{Ground}}
  states and critical currents.
\newblock {\em Physical Review B}, 31(9):5728--5745, May 1985.

\bibitem{Tei88}
S.~Teitel.
\newblock Uniformly frustrated xy models: {{Ground}} state configurations.
\newblock {\em Physica B: Condensed Matter}, 152(1):30--31, August 1988.

\bibitem{LBH+18}
Martijn Lankhorst, Alexander Brinkman, Hans Hilgenkamp, Nicola Poccia, and
  Alexander Golubov.
\newblock Annealed {{Low Energy States}} in {{Frustrated Large Square Josephson
  Junction Arrays}}.
\newblock {\em Condensed Matter}, 3(2):19, June 2018.

\bibitem{BSK09}
Seung~Ki Baek, Hiroyuki Shima, and Beom~Jun Kim.
\newblock Curvature-induced frustration in the \${{XY}}\$ model on hyperbolic
  surfaces.
\newblock {\em Physical Review E}, 79(6):060106, June 2009.

\bibitem{NE18}
Idan Niv and Efi Efrati.
\newblock Geometric frustration and compatibility conditions for
  two-dimensional director fields.
\newblock {\em Soft Matter}, 14(3):424--431, January 2018.

\bibitem{Mey76}
Robert~B. Meyer.
\newblock Structural {{Problems}} in {{Liquid Crystal Physics}}.
\newblock In {Balian R. {and} Weill G.}, editor, {\em Molecular {{Fluids}},
  {{Les Houches Lectures}}, 1973}. {Routledge}, 1976.

\bibitem{HC20}
L.~E. Hough, M.~Spannuth, M.~Nakata, D.~A. Coleman, C.~D. Jones,
  G.~Dantlgraber, C.~Tschierske, J.~Watanabe, E.~K{\"o}rblova, D.~M. Walba,
  J.~E. Maclennan, M.~A. Glaser, and N.~A. Clark.
\newblock Chiral {{Isotropic Liquids}} from {{Achiral Molecules}}.
\newblock {\em Science}, 325(5939):452--456, July 2009.

\bibitem{Sel21}
Jonathan~V. Selinger.
\newblock Director {{Deformations}}, {{Geometric Frustration}}, and {{Modulated
  Phases}} in {{Liquid Crystals}}.
\newblock {\em arXiv:2103.03803 [cond-mat]}, March 2021.

\bibitem{note1}
The form of the Frank free energy was chosen to comply with the quadratic form
  used in the intrinsic approach to frustrated structures \cite{ME21}.

\bibitem{HG21}
Michael~F. Hagan and Gregory~M. Grason.
\newblock Equilibrium mechanisms of self-limiting assembly.
\newblock {\em Reviews of Modern Physics}, 93(2):025008, June 2021.

\end{thebibliography}

\newpage




\begin{figure*}[h]
\includegraphics[width=16.5cm]{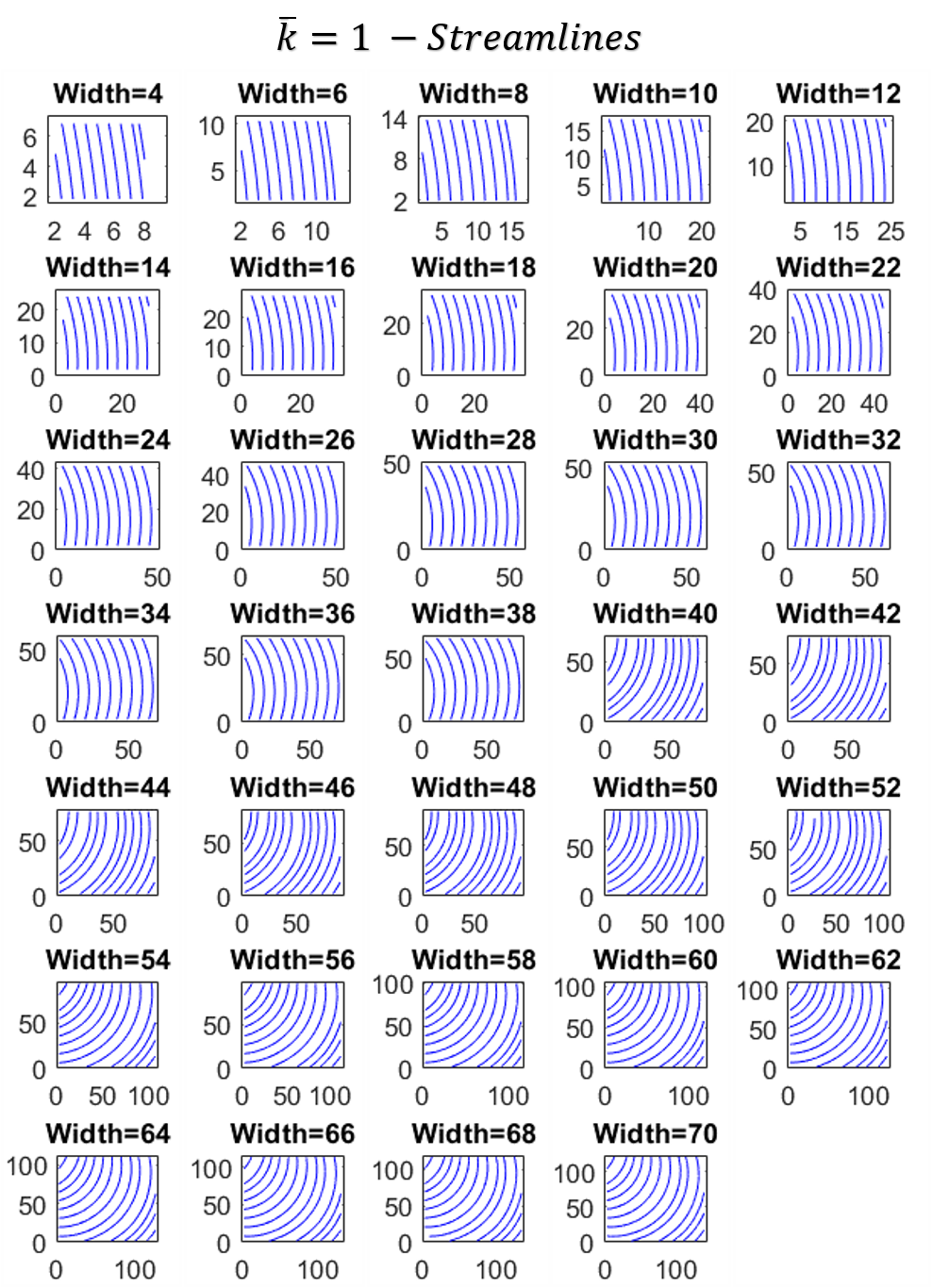}
\caption{ Resulting interpolated streamlines for $\overline{k}=1$. $b_0=0.015$.}
\label{fig:stream1}
\end{figure*}
\newpage

\begin{figure*}[h]
\includegraphics[width=16.5cm]{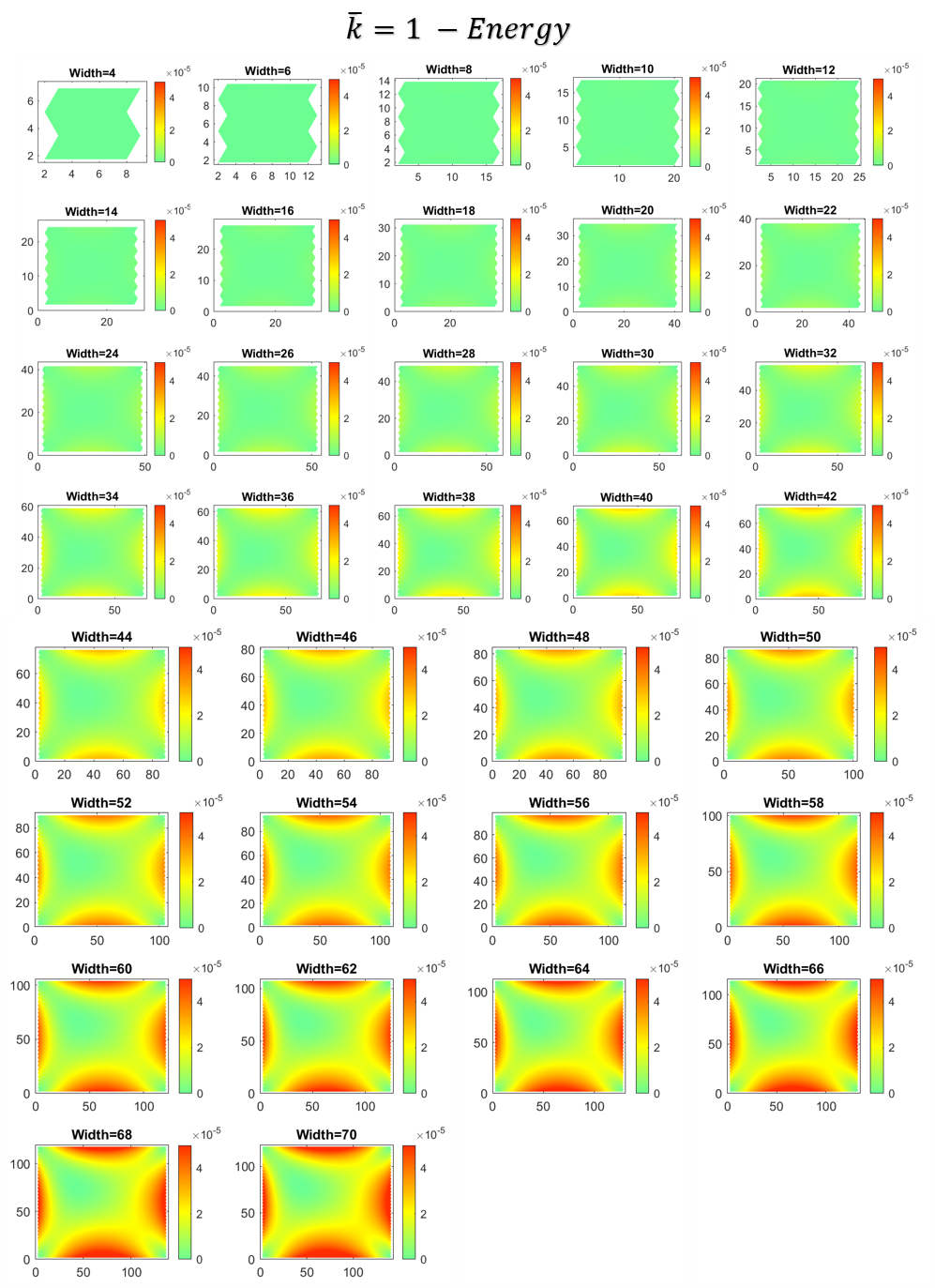}
\caption{ Resulting energy heat-maps for $\overline{k}=1$. $b_0=0.015$.}
\label{fig:ener1}
\end{figure*}

\begin{figure*}[h]
\includegraphics[width=16.5cm]{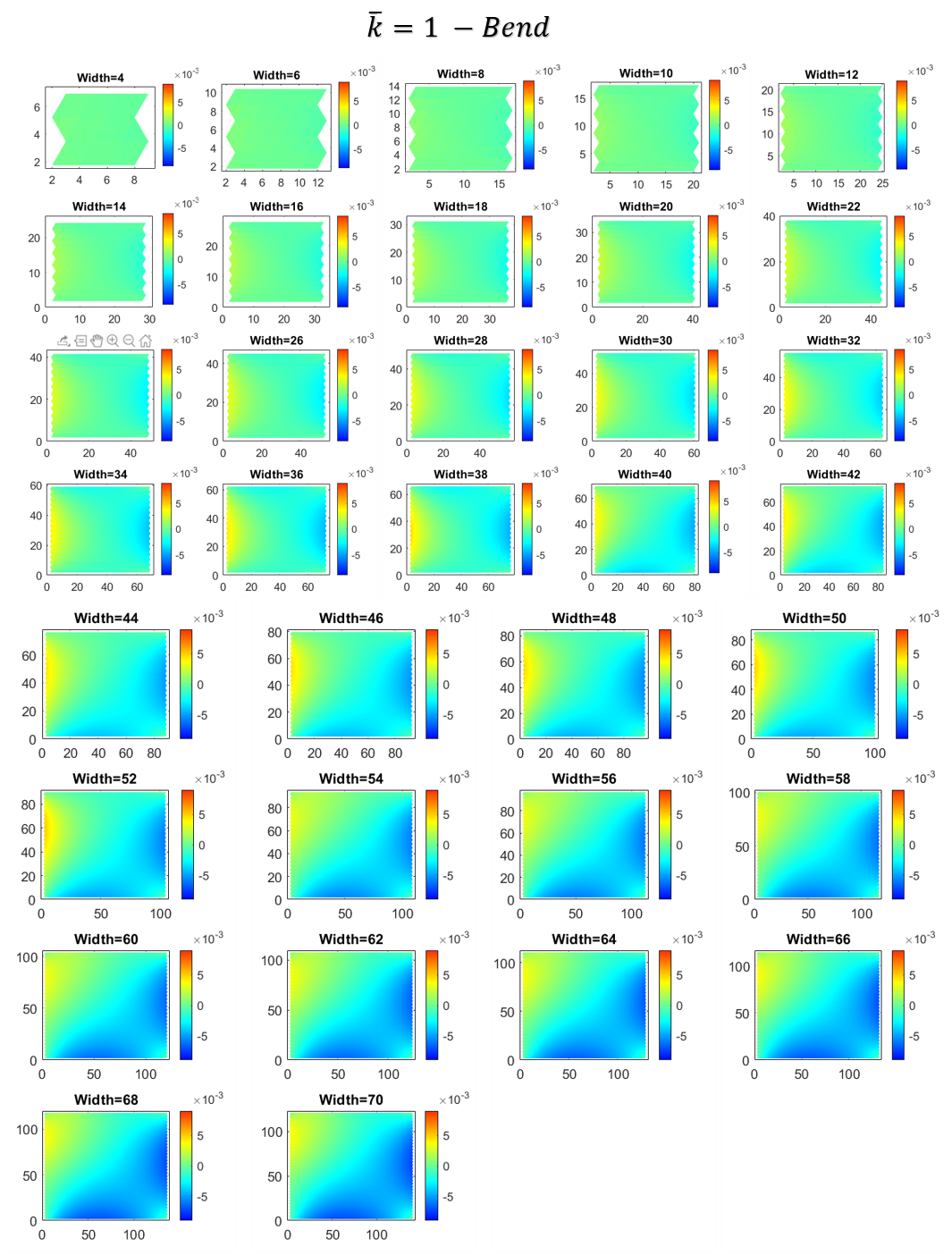}
\caption{ Resulting heat-maps of $b-b_0$ for $\overline{k}=1$. $b_0=0.015$.}
\label{fig:bend1}
\end{figure*}

\begin{figure*}[h]
\includegraphics[width=16.5cm]{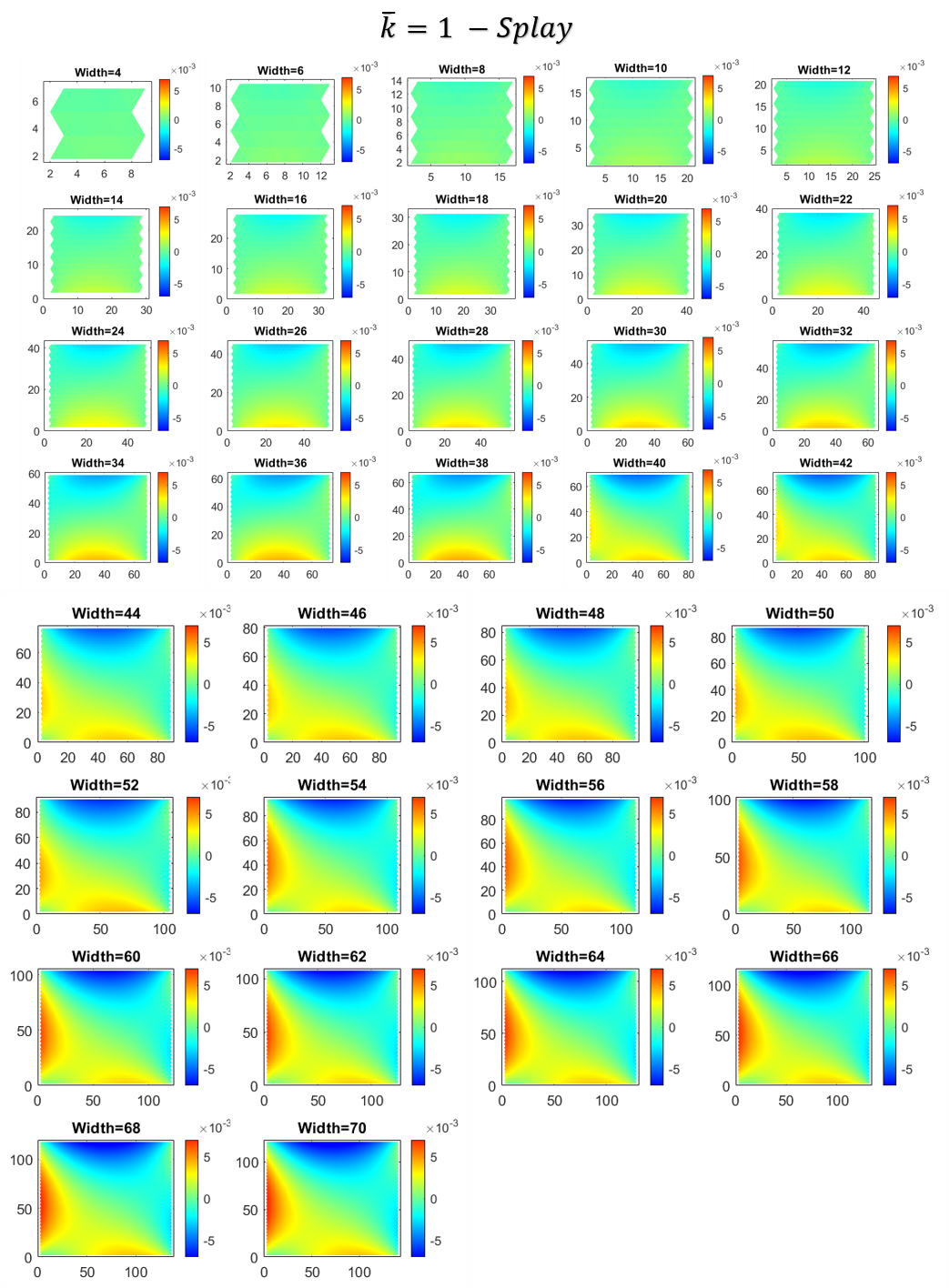}
\caption{ Resulting heat-maps of splay for $\overline{k}=1$. $b_0=0.015$.}
\label{fig:splay1}
\end{figure*}

\newpage
\begin{figure*}[h]
\includegraphics[width=14.5cm]{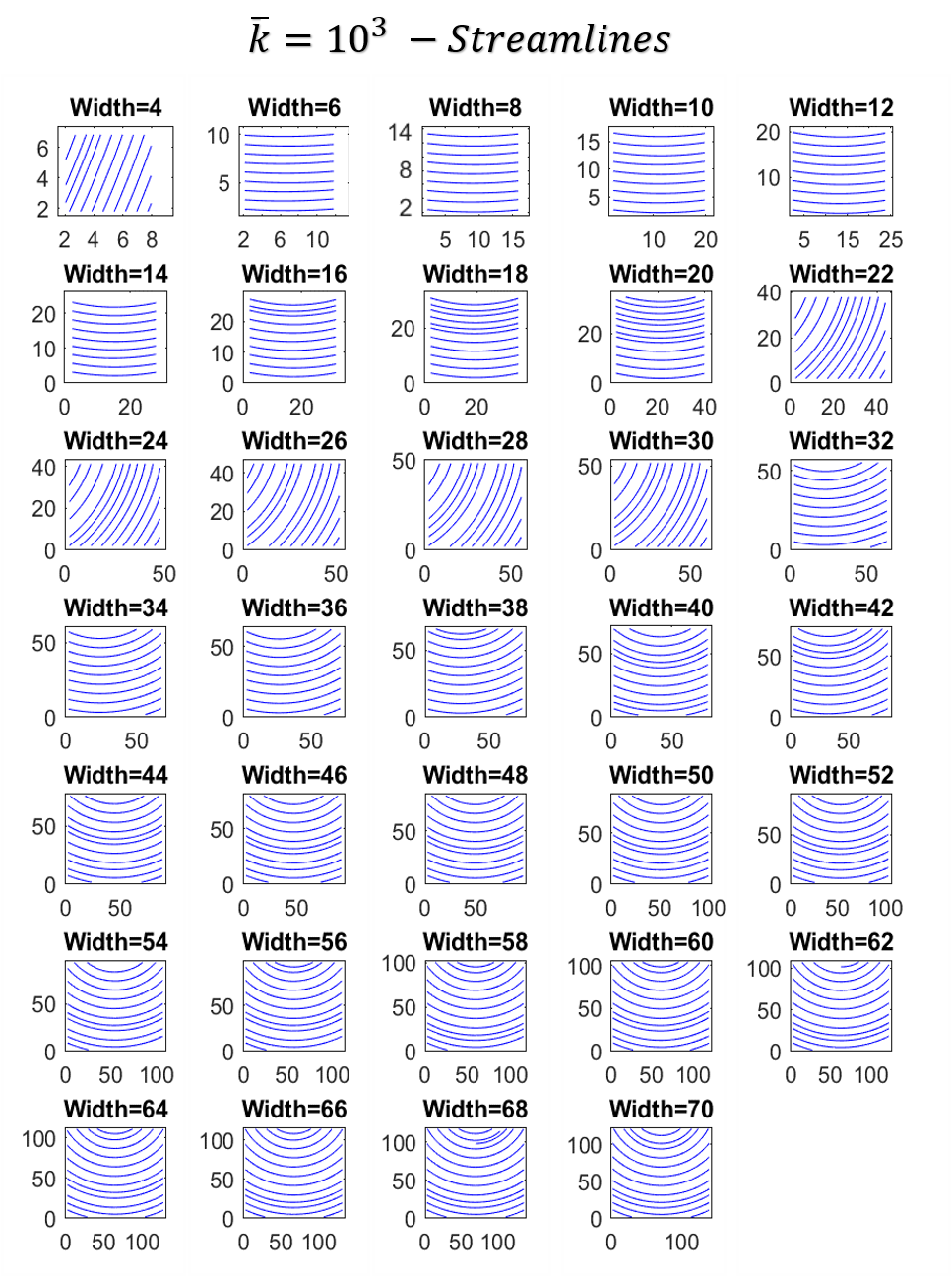}
\caption{ Resulting interpolated streamlines for $\overline{k}=10^3$. $b_0=0.015$.\\
Notice the discontinuous texture orientation jumps of the streamlines between the lattices of size 20 to 22 and between size 30 and 32. The origin of this transition can be attributed to the azimuthal modulation of the energy density visible in Figure 3.(a).}
\label{fig:stream1}
\end{figure*}

\begin{figure*}[h]
\includegraphics[width=16.9cm]{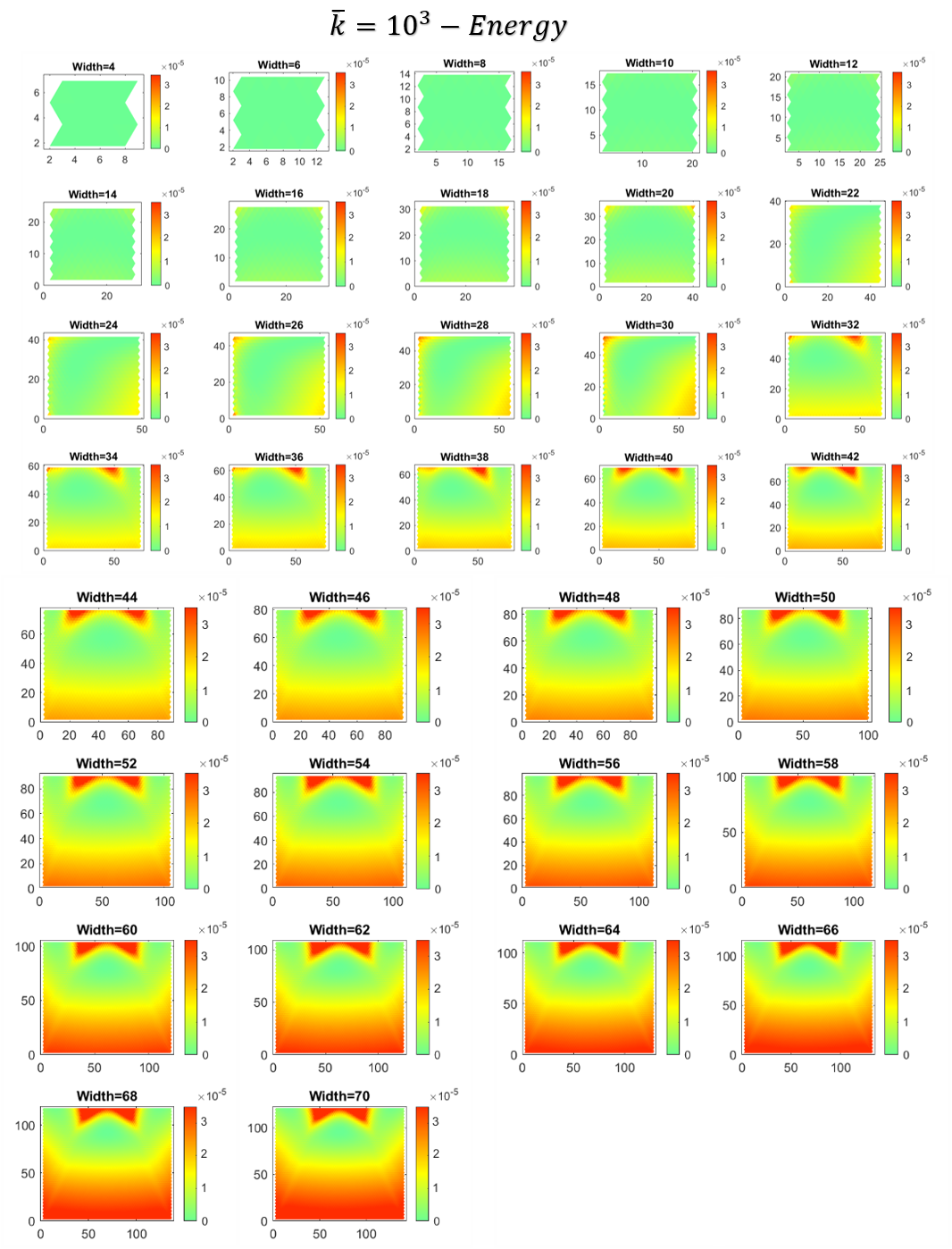}
\caption{ Resulting energy heat-maps for $\overline{k}=10^3$. $b_0=0.015$.}
\label{fig:ener1}
\end{figure*}

\begin{figure*}[h]
\includegraphics[width=16.9cm]{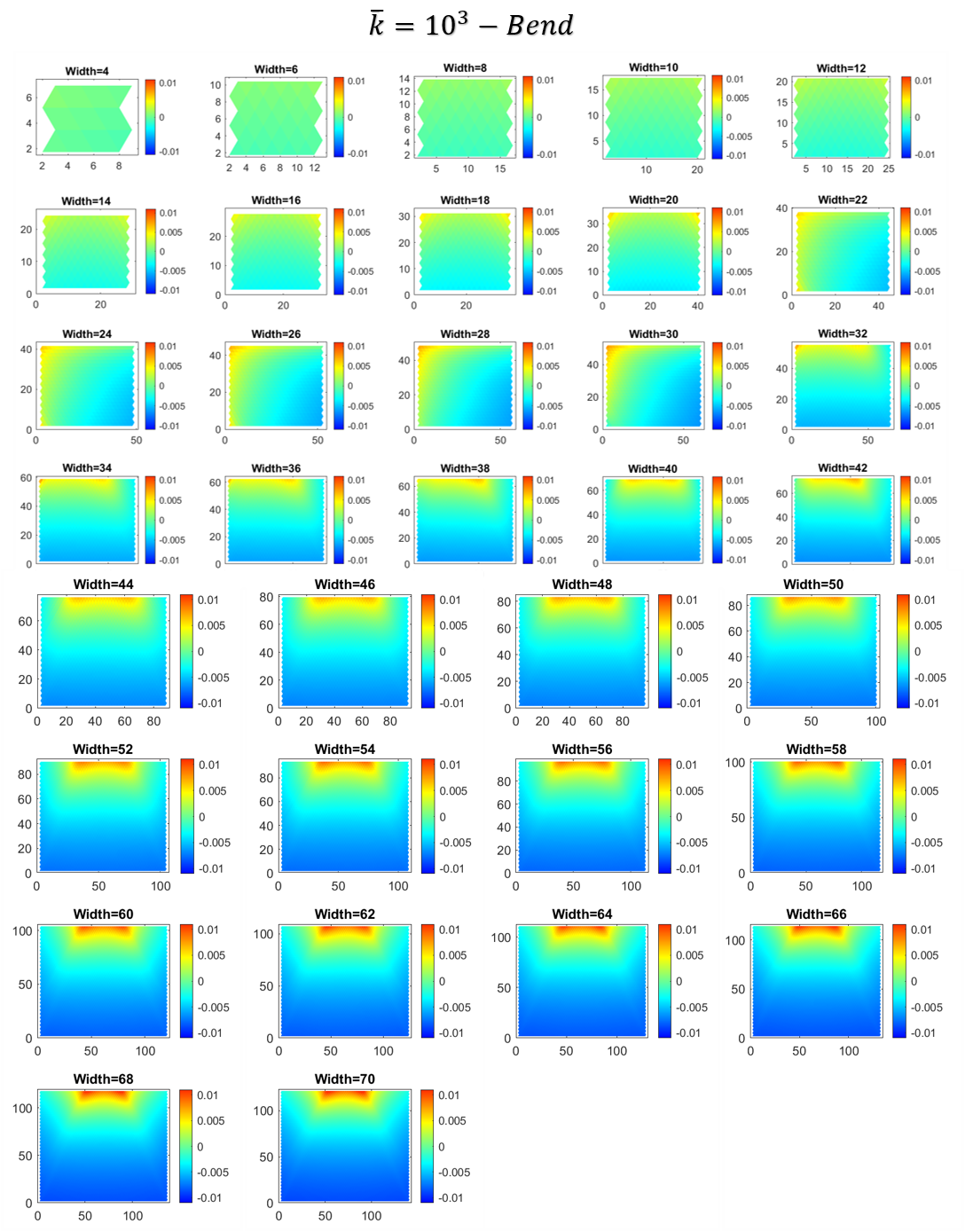}
\caption{ Resulting heat-maps of $b-b_0$ for $\overline{k}=10^3$. $b_0=0.015$.}
\label{fig:bend1}
\end{figure*}

\begin{figure*}[h]
\includegraphics[width=16.9cm]{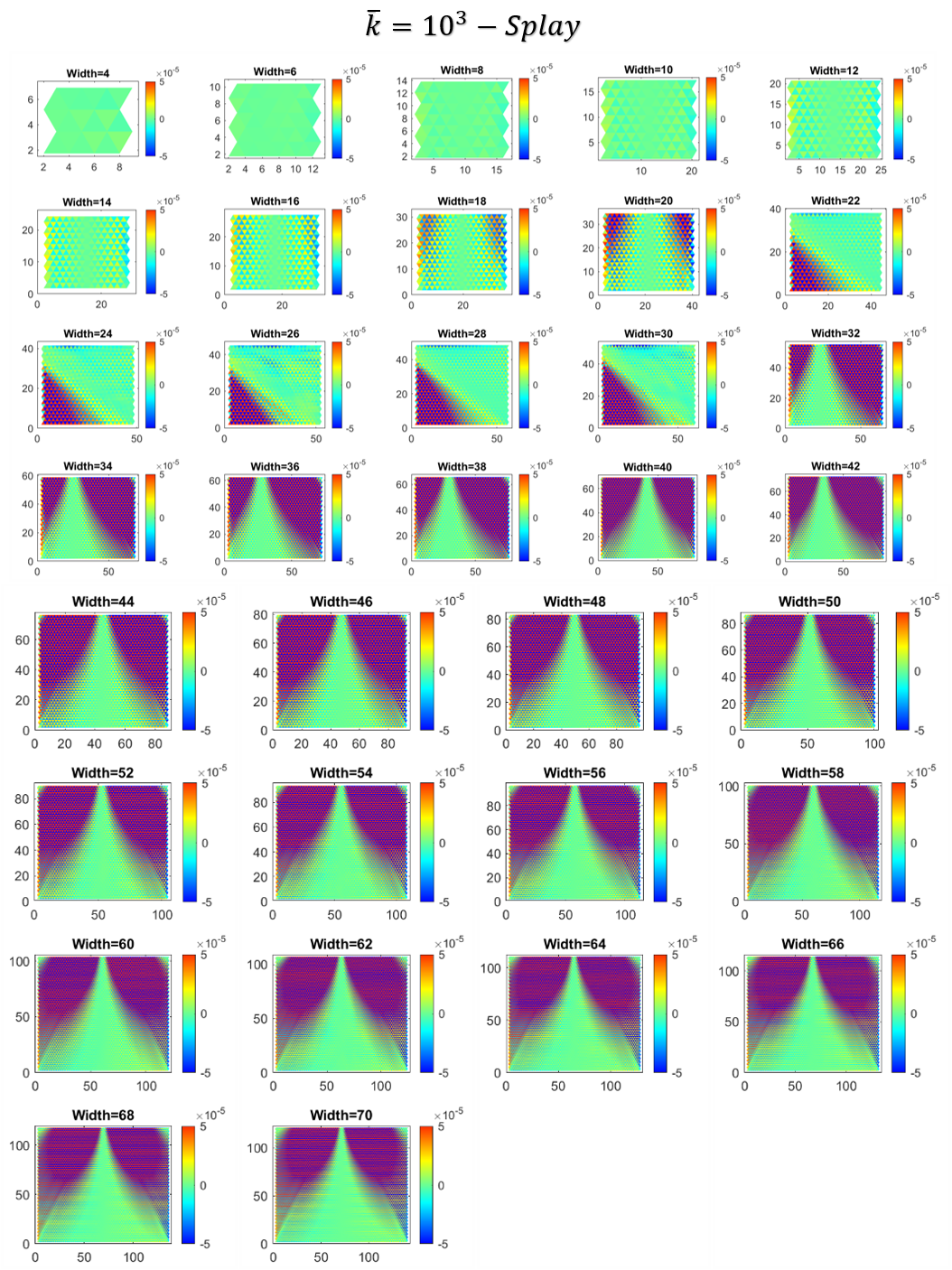}
\caption{ Resulting heat-maps of splay for $\overline{k}=10^3$. $b_0=0.015$.}
\label{fig:splay1}
\end{figure*}

\begin{figure*}[h]
\includegraphics[width=14.5cm]{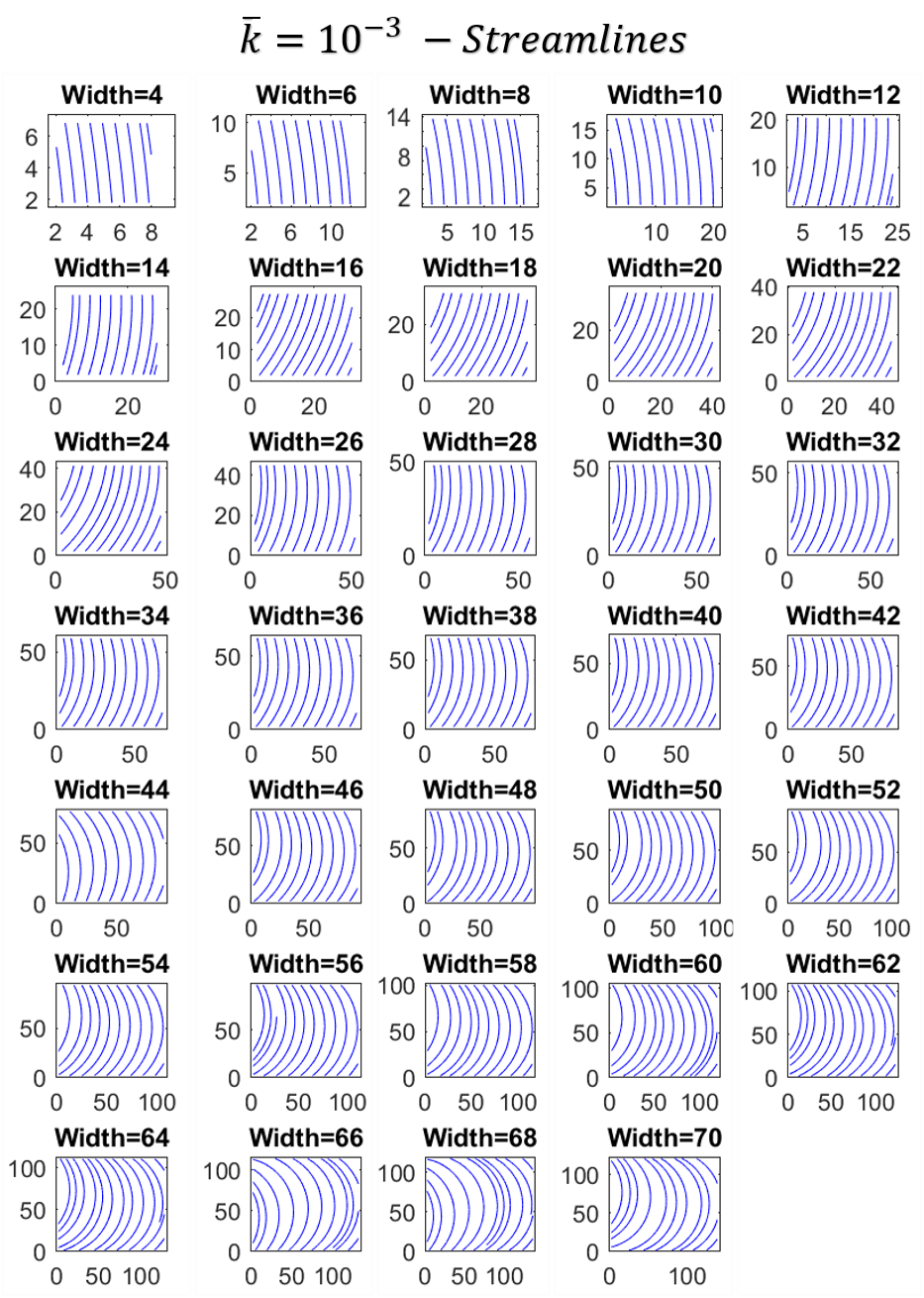}
\caption{ Resulting interpolated streamlines for $\overline{k}=10^{-3}$. $b_0=0.015$.}
\label{fig:stream1}
\end{figure*}

\begin{figure*}[h]
\includegraphics[width=16.9cm]{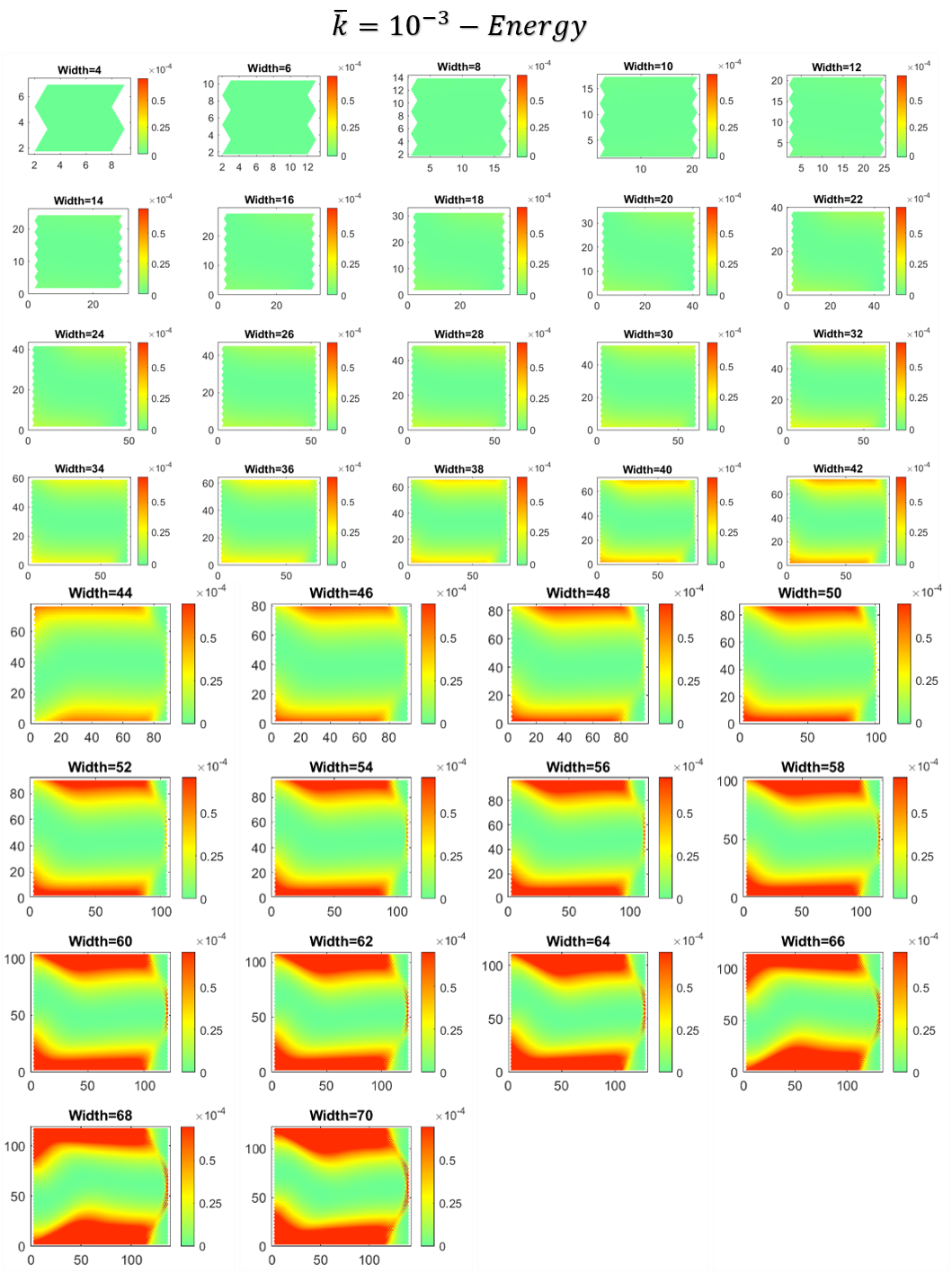}
\caption{ Resulting energy heat-maps for $\overline{k}=10^{-3}$. $b_0=0.015$.}
\label{fig:ener1}
\end{figure*}

\begin{figure*}[h]
\includegraphics[width=16.9cm]{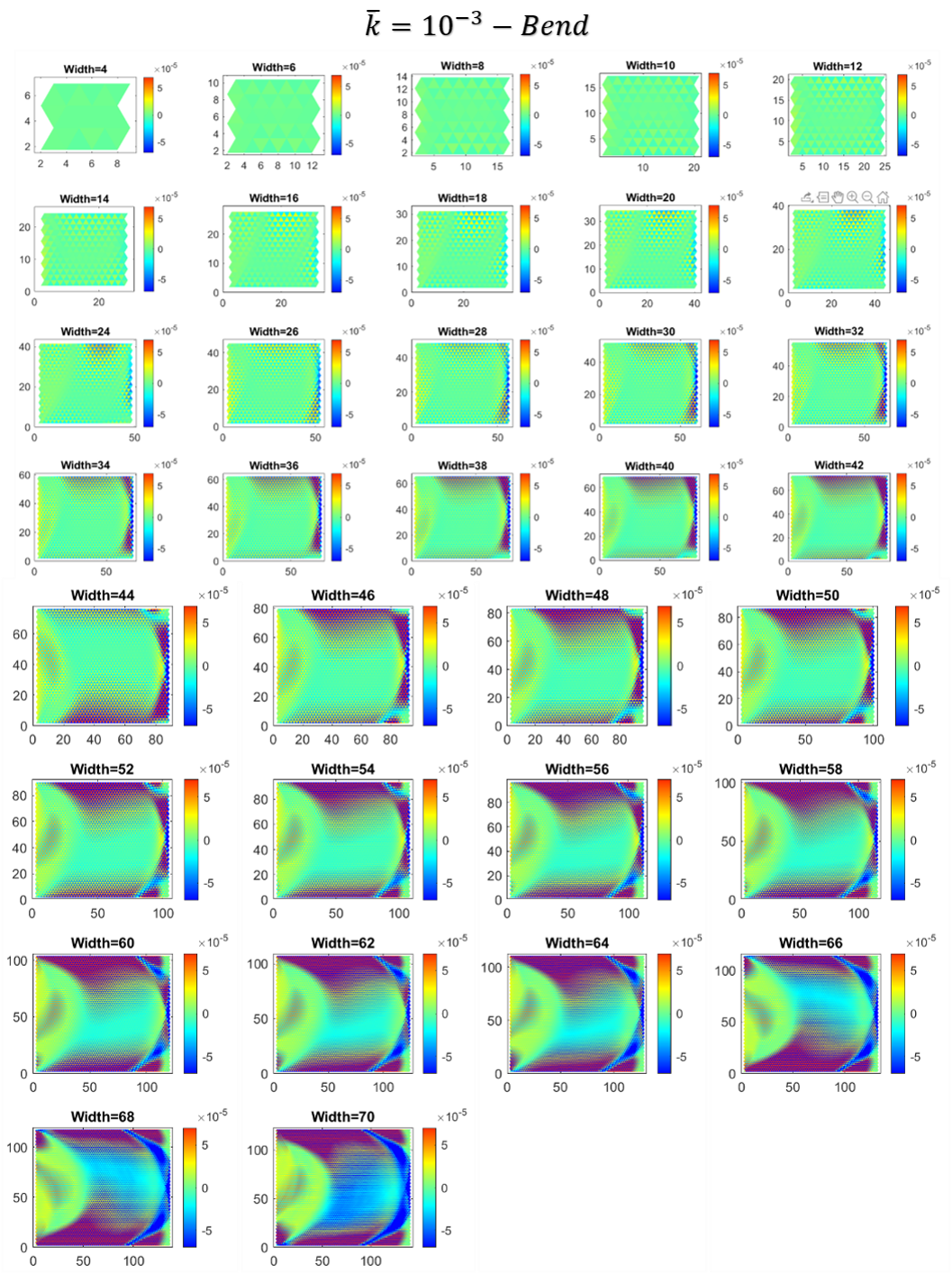}
\caption{ Resulting heat-maps of $b-b_0$ for $\overline{k}=10^{-3}$. $b_0=0.015$.}
\label{fig:bend1}
\end{figure*}

\begin{figure*}[h]
\includegraphics[width=16.9cm]{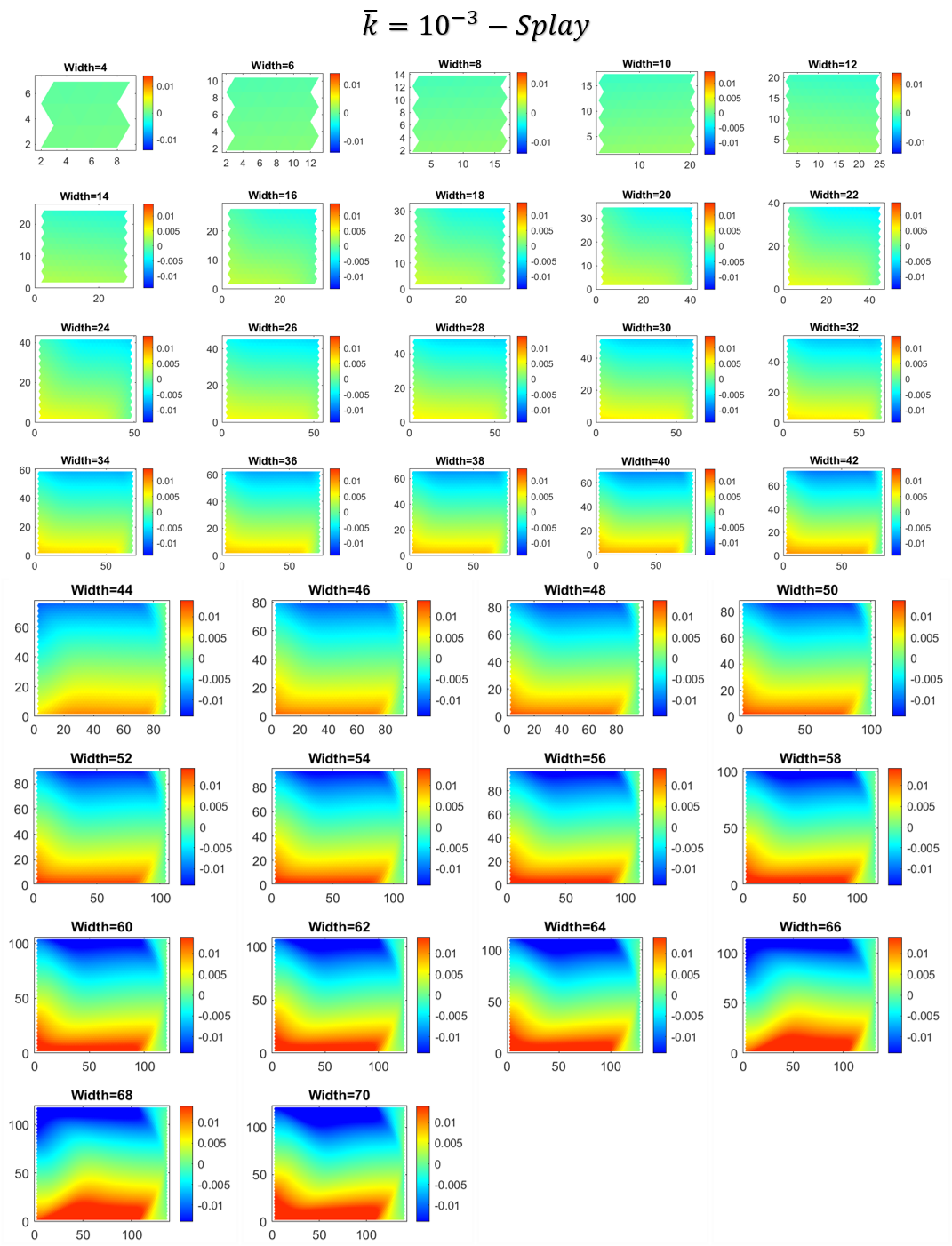}
\caption{ Resulting heat-maps of splay for $\overline{k}=10^{-3}$.}
\label{fig:splay1}
\end{figure*}

\end{document}